\documentclass[
  reprint,
  amsmath,
  amssymb,
  aps,
  pra,
  superscriptaddress,
  longbibliography,
  floatfix
]{revtex4-2}

\usepackage{graphicx}
\usepackage{dcolumn}
\usepackage{bm}
\usepackage{braket}
\usepackage[colorlinks=true,linkcolor=blue,urlcolor=black,citecolor=blue,bookmarksopen=true]{hyperref}
\usepackage{xcolor}
\usepackage{mathtools}
\usepackage{booktabs}
\usepackage{multirow}
\usepackage{physics}
\usepackage[caption=false]{subfig}
\hypersetup{hypertexnames=false}

\newcommand{\pcsdpnl}{Physical and Computational Sciences Division, Pacific Northwest National Laboratory, Richland, WA, 99354, USA}
\newcommand{\eedpnl}{Energy and Environment Division, Pacific Northwest National Laboratory, Richland, WA, 99354, USA}

\newcommand{\ubc}{Department of Electrical and Computer Engineering, the University of British Columbia, Vancouver, BC V6T 1Z4, Canada}

\newcommand{\mean}[1]{\langle #1 \rangle}
\newcommand{\R}{\mathbb{R}}
\newcommand{\dt}{\Delta t}

\newcommand{\diag}{\mathrm{diag}}
\newcommand{\onehot}{\mathrm{one\mbox{-}hot}}
\newcommand{\binary}{\mathrm{binary}}

\newcommand{\Sch}{Schr\"odinger}

\begin{document}

\title{Encoding Choices and Fault-Tolerant Resource Estimates for Digital Quantum Hamiltonian Descent}

\author{Chenxu Liu}
\email{chenxu.liu@pnnl.gov}
\affiliation{\pcsdpnl}

\author{Meng Wang}
\affiliation{\ubc}

\author{Mingze Li}
\affiliation{\eedpnl}

\author{Muqing Zheng}
\affiliation{\pcsdpnl}

\author{Samuel Stein}
\affiliation{\pcsdpnl}

\author{Yousu Chen}
\affiliation{\eedpnl}

\date{\today}

\begin{abstract}
Quantum Hamiltonian descent (QHD) formulates continuous optimization as time-dependent quantum dynamics, where a kinetic term drives exploration and a potential term encodes the objective function. Digital implementations of QHD require encoding the search space into qubits, and this choice can shift the dominant cost among logical qubits, circuit depth, non-Clifford rotations, and potential synthesis. In this work, we present an encoding-aware resource analysis comparing one-hot and binary amplitude encodings for QHD. We derive gate-count scalings, construct and validate circuits against classical \Sch-equation solvers, and estimate Clifford+$R_z$ and fault-tolerant Clifford+$T$ resources on benchmark optimization problems. Binary encoding reduces the data register from $O(dN)$ to $O(d\log N)$ qubits and gives comparable asymptotic scaling for both kinetic and potential evolutions. Across all benchmark problems studied, binary encoding also uses fewer $R_z$ rotations than one-hot encoding, making it the preferred option for fault-tolerant implementations where arbitrary rotations dominate the cost. 
Kinetic approximations based on low-momentum spectra and approximate QFTs can further reduce the binary kinetic cost to polylogarithmic scaling. However, for targets such as Ackley, potential synthesis can dominate the total cost and reduce the benefit of kinetic approximations. These results suggest that exploiting the analytic structure of the target function to compile the potential evolution in QHD more efficiently is needed for further resource reductions.
\end{abstract}

\maketitle

\section{Introduction}
\label{sec:introduction}

By taking advantage of quantum-mechanical principles, quantum computers are expected to solve selected problems substantially faster than classical computers. This possibility has motivated quantum algorithms for cryptography~\cite{shor1999polynomial}, optimization~\cite{shor1994algorithms, Zhou2020_qaoa}, quantum simulation and chemistry~\cite{georgescu2014quantum, kandala2017hardware}, finance~\cite{rebentrost2018quantum, woerner2019quantum}, and other computationally demanding applications. With rapid progress in quantum hardware, the field is beginning to move beyond the noisy intermediate-scale quantum (NISQ) regime toward fault-tolerant quantum computation~\cite{NISQ_Preskill, Megaquop}. Fault-tolerant quantum computers will enable addressing large-scale, classically challenging problems across these application areas. This makes it important to quantify the resources required by candidate quantum algorithms and to identify implementation strategies that can make them practical on future quantum computers.

Optimization is a particularly important target because optimization problems arise throughout physics and chemistry~\cite{georgescu2014quantum, kandala2017hardware}, logistics and engineering design~\cite{lavalle2006planning, betts2010practical, nocedal1999numerical}, finance~\cite{markowitz1952portfolio, cornuejols2007optimization}, machine learning~\cite{zafar2017fairness, tibshirani1996regression, xu2009robustness}, and power systems~\cite{low2014convex, aravena2023recent, holzer2024grid}. Existing quantum approaches to optimization include algorithms that accelerate algebraic subroutines of classical optimization methods, such as linear-system or interior-point steps, and Hamiltonian-based approaches that use quantum dynamics to search for low-energy configurations~\cite{wu2025quantumdual, Zhou2020_qaoa, Das2008_annealing, abbas2024challenges}. Representative Hamiltonian-based methods include quantum annealing, adiabatic quantum optimization, and the quantum approximate optimization algorithm (QAOA)~\cite{Das2008_annealing, Tasseff2024_annealing, kim2025quantum, Zhou2020_qaoa}. These methods are most naturally formulated for discrete variables, where the objective is encoded as an Ising, QUBO, or related spin Hamiltonian. Applying them to continuous-variable optimization typically requires discretizing each variable and encoding the resulting grid values into binary, one-hot, domain-wall, or related representations, which introduces a resolution-accuracy tradeoff and can obscure the smooth structure of the original objective.

Recently, quantum Hamiltonian descent (QHD) was proposed as a quantum-dynamical algorithm natively designed for continuous-variable optimization~\cite{leng2023quantum, kushnir2025qhdopt}. In QHD, the objective function is encoded as the potential energy of an effective particle, while a time-dependent kinetic term controls exploration of the search domain. The wavefunction initially explores broadly and is then driven toward localization near low-potential regions. Previous studies have applied QHD to unconstrained continuous optimization problems and observed promising behavior for finding global minima~\cite{leng2023quantum, kushnir2025qhdopt}. Extensions based on augmented Lagrangian ideas have also been discussed for constrained optimization~\cite{wu2026ALQHD, li2025QHD}. However, noisy quantum hardware can limit the reliability of optimization outputs, and accurate implementations for nontrivial instances may require systematic error control or quantum error correction~\cite{abbas2024challenges, wu2026ALQHD}. This motivates studying QHD in the digital fault-tolerant model, where qubits, gates, non-Clifford resources, and space-time volume can be estimated explicitly.

A central issue in a digital implementation of QHD is how the discretized continuous domain is embedded into qubits. Different encodings can represent the same grid with very different circuit structures. 
Recent work by Wu {\it et al.} applied QHD to constrained optimization using an augmented-Lagrangian formulation and estimated the corresponding fault-tolerant (FT) resources with a coordinate-wise one-hot encoding~\cite{wu2026ALQHD}. The analysis in Ref.~\cite{wu2026ALQHD} provided an important application-level resource estimate for QHD, but it left open how the encoding choice itself affects the resources required for digital FT implementations.
A coordinate-wise one-hot encoding uses $O(dN)$ qubits for $d$ variables and $N$ grid points per variable, but it maps the finite-difference kinetic operator to local nearest-neighbor hopping terms. In contrast, a binary amplitude encoding uses only $O(d\log N)$ qubits, but the kinetic operator becomes a sparse transition operator over binary strings and the diagonal potential can require denser phase synthesis with high-weight Pauli-Z rotations. However, the resulting tradeoff between qubit count, circuit depth, non-Clifford cost, and surface-code space-time volume has not been systematically explored. Understanding this encoding dependence is therefore a central step toward developing concrete implementation plans for QHD on FT quantum computers.

In this paper, we develop an encoding-aware resource analysis for QHD applied to digital FT quantum computers that connects analytic gate-count scaling, explicit circuit construction, numerical validation, and FT resource estimates. Building on the FT perspective introduced in Ref.~\cite{wu2026ALQHD}, we shift the focus to the encoding dependence of the same discretized QHD dynamics. 
Specifically, we compare coordinate-wise one-hot encoding with binary amplitude encoding, derive the kinetic- and potential-evolution costs for each representation, and implement the corresponding Clifford+$R_z$ circuits. We then evaluate how logical-qubit requirements, $R_z$ rotation counts, and Clifford+$T$ synthesis costs change across representative benchmark objectives.
The scaling analysis shows that binary encoding reduces the data register from $O(dN)$ to $O(d\log N)$ qubits while giving comparable asymptotic $R_z$-rotation scaling for both kinetic and potential evolutions. In all benchmark problems, binary encoding also uses fewer $R_z$ rotations than one-hot encoding. Since arbitrary rotations become costly non-Clifford resources after Clifford+$T$ synthesis, this makes binary encoding the preferred option for FT implementations of QHD under the assumed resource model. 

We also consider approximated implementations that can be adopted with binary encoding, including approximate-QFT (AQFT) and low-momentum kinetic-spectrum approximations, which can further reduce the binary kinetic cost to polylogarithmic scaling in the grid resolution. However, when the target function is difficult to compile, as in dense transcendental objectives such as Ackley, potential synthesis can dominate the total cost and reduce the practical benefit of kinetic approximations. Our results suggest that further resource reductions for digital QHD should focus on exploiting the analytic structure of the objective function to compile the potential evolution more efficiently.

The rest of the paper is organized as follows. Sec.~\ref{sec:qhd} reviews the QHD algorithmic framework. Sec.~\ref{sec:one-hot} presents the one-hot encoding, its circuit implementation, and the corresponding gate-count scaling analysis. Sec.~\ref{sec:binary} develops the binary-encoding implementation, including exact and approximate kinetic-evolution constructions, and, specifically, Sec.~\ref{sec:binary:comparison} compares the resulting scaling with the one-hot approach. Sec.~\ref{sec:circuits} describes the circuit realization and numerical validation of the QHD dynamics. Sec.~\ref{sec:resource} reports benchmark gate counts and FT resource estimates. Sec.~\ref{sec:discussion} discusses implications and limitations, and Sec.~\ref{sec:conclusion} concludes the paper.

\section{Quantum Hamiltonian Descent}
\label{sec:qhd}
Quantum Hamiltonian descent (QHD) is a quantum-dynamical algorithm for continuous optimization. It maps the optimization problem to the dynamics of an effective particle in a high-dimensional space, with the objective function encoded as the particle's potential energy. A time-dependent kinetic energy, equivalent to an effective particle mass, controls how broadly the wavefunction explores the search domain. The dynamics are designed so that the particle initially has enough mobility to explore many configurations and is gradually driven toward localization in low-potential regions, ideally concentrating probability near the global minimum. 

We first briefly review the mathematical construction used in this work. A more detailed discussion can be found in Refs.~\cite{leng2023quantum, kushnir2025qhdopt}. Consider an optimization problem with target function $f(\mathbf{x})$ over a bounded domain $\mathbf{x} \in \Omega \subset \R^d$. QHD maps this problem to an effective particle whose wavefunction evolves under a time-dependent Hamiltonian of the form
\begin{align}
  H(t) & = H_K(t) + H_V(t), \nonumber \\
  & = e^{\phi(t)}\left(-\frac{\nabla^2}{2}\right) + e^{\chi(t)} f(\mathbf{x}),
  \label{eq:qhd_hamiltonian}
\end{align}
where the kinetic term ($H_K$) enables delocalized exploration and the potential term ($H_V$) encodes the objective landscape. To make the mapping clearer, we can define an effective mass $m_{\text{eff}} = e^{\chi(t) - \phi(t)}$, so that the rescaled Hamiltonian becomes $\tilde{H}(t) = -\nabla^2/(2m_{\text{eff}}) + f(\mathbf{x})$. In Ref.~\cite{leng2023quantum}, one choice of the time-dependent functions is
\begin{align}
  e^{\phi(t)} = \frac{1}{1+\gamma t^2}, \quad e^{\chi(t)} = 1+\gamma t^2,
\end{align}
so that the effective mass becomes $m_{\text{eff}} = (1 + \gamma t^2)^2$. The resulting dynamics evolves from a light-particle regime to a heavy-particle regime, which causes the wavefunction to become increasingly localized.

As pointed out in Ref.~\cite{leng2023quantum}, QHD dynamics can be simulated on a digital quantum computer by discretizing the search domain and Trotterizing the resulting finite-dimensional Hamiltonian. For each coordinate, we choose grid points $x_{\alpha, j}$ with spacing $h_{\alpha}$, where $\alpha \in \{1, 2, \dots, d\}$ labels the search-space dimension and $j \in \{1, 2, \dots, N_\alpha\}$ labels the corresponding grid point. A real-space basis state is then labeled by a multi-index $\mathbf{j}=(j_1,\ldots,j_d)$ and corresponds to the grid point $\mathbf{x}_{\mathbf{j}}=(x_{1,j_1},\ldots,x_{d,j_d})$. The wavefunction on the discretized domain can be represented as
\begin{align}
  \ket{\psi}
  & =
  \sum_{j_1=1}^{N_1}\cdots\sum_{j_d=1}^{N_d}
  \psi(\mathbf{x}_{\mathbf{j}})
  \ket{\mathbf{x}_{\mathbf{j}}}, \nonumber \\
  \ket{\mathbf{x}_{\mathbf{j}}}
  & =
  \ket{x_{1,j_1}}\otimes\cdots\otimes\ket{x_{d,j_d}}.
\end{align}
On this grid, the potential term is diagonal in the real-space basis,
\begin{equation}
  H_V(t)=e^{\chi(t)}D_f,
  \ 
  D_f=\diag\{f(x_{1,j_1},\ldots,x_{d,j_d})\}_{\mathbf{j}}.
\end{equation}
If the target function can be decomposed into terms with small variable support, then each term can be implemented only on the Hilbert space associated with its supported variables, rather than by constructing a diagonal operator over all qubits. 

With the spatial discretization, the kinetic term is obtained by replacing the Laplacian operator with the finite-difference approximation. In one dimension, for example, the standard second-difference operator gives
\begin{equation}
  L_h=\frac{1}{h^2}
  \begin{pmatrix}
    -2 & 1 & 0 & \cdots \\
    1 & -2 & 1 & \cdots \\
    0 & 1 & -2 & \cdots \\
    \vdots & \vdots & \vdots & \ddots
  \end{pmatrix},
  \ 
  H_K(t)=-\frac{e^{\phi(t)}}{2}L_h,
  \label{eq:laplacian}
\end{equation}
with the multidimensional kinetic operator formed by the corresponding Kronecker sum over coordinates.  

Thus, an exact digital representation of the discretized dynamics is obtained by implementing exponentials of the finite-dimensional matrices $H_V(t)$ and $H_K(t)$. 
As an example, a first-order Trotter step for Eq.~\eqref{eq:qhd_hamiltonian} at midpoint time $t_s= s \dt$ can be computed as
\begin{equation}
  U_s \approx
  e^{-i \dt H_K(t_s)}
  e^{-i \dt H_V(t_s)}.
  \label{eq:first_order_trotter}
\end{equation}

Ref.~\cite{leng2023quantum} also noted that Hamiltonian embedding~\cite{leng2025hamiltonianembedding} can use additional qubits to simplify the implementation of the kinetic and potential terms, thereby reducing circuit complexity and making QHD more suitable for near-term quantum devices. In the next section, we discuss how QHD simulation can be realized using one-hot and binary amplitude encodings.

\section{QHD with One-hot Encoding Scaling Analysis} \label{sec:one-hot}

In this section, we discuss QHD implementations based on one-hot encoding~\cite{leng2023quantum, kushnir2024qhdopt, wu2026ALQHD} and estimate their complexity for FT quantum hardware. Unless otherwise stated, we assume that the optimization problem has $d$ independent variables, each discretized into $N$ grid points with grid spacing $h$, where $N = 2^n$ and $n \in \mathbb{Z}_{+}$. We focus primarily on periodic boundary conditions; cases with Dirichlet boundary conditions are specified explicitly.

In the rest of this section, we first state the assumptions underlying the FT resource model in Sec.~\ref{sec:one-hot:res-model}. We then review the one-hot QHD encoding in Sec.~\ref{sec:one-hot:encoding} and analyze the rotation-count scaling of the kinetic and potential terms in Sec.~\ref{sec:one-hot:kinetic-count} and~\ref{sec:one-hot:potential-count}, respectively. The resulting Clifford+$T$ cost is analyzed in Sec.~\ref{sec:one-hot:t-count}, and the main scaling results are summarized in Sec.~\ref{sec:one-hot:summary}.

\subsection{Fault-Tolerant Resource Model} \label{sec:one-hot:res-model}

In the rest of the paper, we analyze the resource scaling of the one-hot and binary amplitude implementations of QHD in the fault-tolerant quantum computing (FTQC) regime and make scaling comparisons. As discussed in Ref.~\cite{wu2026ALQHD}, practical applications of QHD may require quantum error correction (QEC) to control accumulated algorithmic and hardware errors, so the relevant resource model should be formulated at the logical level. We use the surface code as a representative FT architecture. In such architectures, Clifford operations, including Pauli gates, CNOT gates, and multi-qubit Pauli measurements implemented through lattice-surgery-type operations, are usually treated as comparatively low-cost logical operations, although they still consume logical qubit patches and code cycles. 

The dominant cost instead comes from non-Clifford operations. In particular, arbitrary phase rotations such as $R_z(\theta)$ are not generally available as native protected logical gates. By the Eastin-Knill theorem, a QEC code cannot provide a universal set of transversal logical gates, and continuous rotation families must therefore be implemented through additional FT constructions rather than assumed to be free logical operations~\cite{Eastin2009}. In surface-code-based FTQC, such rotations are typically approximated by Clifford+$T$ circuits, with the required $T$ states supplied through magic-state distillation or related protocols~\cite{Bombin2009,Horsman2012,Brown2017,Litinski2019,litinski2019magic,gidney2024cultivation,rosenfeld2025,Laflamme2014,Anderson2014}. This separation between Clifford and non-Clifford resources is further emphasized in Pauli-based compilation (PBC), where Clifford operations can be propagated, absorbed, or removed from the explicit computation, leaving the logical workload organized primarily in terms of non-Clifford Pauli rotations and Pauli measurements~\cite{Bravyi2016, Litinski2019}. Therefore, the leading resource cost is governed not by the number of entangling Clifford gates as in NISQ devices, but by the number and precision of non-Clifford rotations, especially the $T$ gates generated when synthesizing the $R_z$ and multi-qubit Pauli rotations appearing in the QHD implementation. We therefore formulate the following scaling analysis in terms of logical qubit counts, Pauli-rotation counts, synthesis precision, and the resulting $T$-gate and magic-state overhead.

\subsection{QHD with One-hot encoding}
\label{sec:one-hot:encoding}

In the one-hot encoding, each variable $x_\alpha$ is represented by an $N$-qubit register. The $j$-th grid point $x_{\alpha,j}$ is encoded as
\begin{equation}
  \ket{j} \mapsto \ket{0\cdots010\cdots0},
\end{equation}
where the only occupied qubit is the $j$-th qubit of the corresponding $N$-qubit register. Therefore, the total number of logical data qubits required to encode the full QHD simulation is
\begin{equation}
  n_{\onehot}=d \cdot N .
  \label{eq:one-hot-nq}
\end{equation}
If auxiliary qubits are used for specific synthesis routines, the logical qubit count becomes
\begin{equation}
  n_{\mathrm{logical}} = d \cdot N+n_{\mathrm{anc}} .
\end{equation}
For the baseline implementation considered in this paper, the kinetic and potential terms can be compiled using Pauli rotations and parity networks without requiring a large number of additional auxiliary logical qubits.

With the one-hot encoding, the kinetic and potential operators acquire effective low-weight Pauli representations. In particular, operators on the spatial basis states can be written as~\cite{leng2023quantum},
\begin{align}
    \dyad{j} 
    &\mapsto \frac{I-Z_j}{2},  \\
    \dyad{j}{j+1}+\dyad{j+1}{j} 
    &\mapsto \frac{1}{2}\left(X_jX_{j+1}+Y_jY_{j+1}\right),
\end{align}
where $X_j$, $Y_j$, and $Z_j$ are Pauli operators acting on the $j$-th qubit of the one-hot register for the corresponding variable. We now use these representations to count the number of rotations required for one QHD Trotter step.

\subsection{Kinetic term rotation gate count}
\label{sec:one-hot:kinetic-count}

The finite-difference kinetic operator maps to Pauli strings on neighboring qubits. For a single one-hot register, the kinetic Hamiltonian takes the form
\begin{align}
  H_K(t)
  =
  -\frac{e^{\phi(t)}}{4h^2}
  \bigg[
    &\sum_{j=0}^{N-2}
      \left( X_jX_{j+1}+Y_jY_{j+1} \right) 
      \nonumber\\
    &+X_{N-1}X_0+Y_{N-1}Y_0
  \bigg],
\end{align}
where the final term enforces periodic boundary conditions.

The one-hot kinetic term can be implemented directly in the position basis. The nearest-neighbor Pauli strings can be grouped into two commuting layers, corresponding to bonds beginning on even and odd sites. A first-order even-odd decomposition gives
\begin{align}
  U_K(t)
  &=
  e^{-iH_K(t)\Delta t} \nonumber \\
  &\approx
  \prod_{\substack{j=0 \\ j\,\mathrm{even}}}^{N-2}
  \exp\!\left[
    i\frac{\Delta t\, e^{\phi(t)}}{4h^2}
    \left(X_jX_{j+1}+Y_jY_{j+1}\right)
  \right]
  \nonumber\\
  &\quad\times
  \prod_{\substack{j=1 \\ j\,\mathrm{odd}}}^{N-1}
  \exp\!\left[
    i\frac{\Delta t\, e^{\phi(t)}}{4h^2}
    \left(X_jX_{j+1}+Y_jY_{j+1}\right)
  \right],
  \label{eq:onehot_even_odd_kinetic}
\end{align}
where $j+1$ is understood modulo $N$.

For resource counting, each variable contributes $N$ nearest-neighbor bonds. Each bond contains one $XX$ rotation and one $YY$ rotation. Thus, the number of kinetic Pauli rotations per kinetic layer is
\begin{equation}
    M_K = 2dN = O(dN).
    \label{eq:onehot-MK}
\end{equation}
For $N_t$ QHD Trotter steps, the total number of kinetic rotations is
\begin{equation}
    N^{K}_{\mathrm{rot}}
    =
    N_t M_K
    =
    2N_t dN .
    \label{eq:onehot-NrotK}
\end{equation}

Each $XX$ or $YY$ rotation can be compiled into a constant-size Clifford circuit together with one arbitrary single-qubit $R_z$-type rotation. Consequently, the Clifford and CNOT overhead of the kinetic part scales linearly with the number of kinetic rotations,
\begin{equation}
    N_{\mathrm{CX}}^{K}
    =
    O(N_tM_K)
    =
    O(N_t dN).
\end{equation}

\subsection{Potential term rotation gate count}
\label{sec:one-hot:potential-count}

The potential term is diagonal in the position basis and can be expressed using one-hot occupation operators on each qubit. Consider first a potential contribution that depends only on a single variable, $g(x_\alpha)$. Representing this function on the spatial grid gives,
\begin{align}
    g(x_\alpha)
    &\mapsto 
    \sum_{j=0}^{N-1} g(x_{\alpha,j}) \dyad{j},
    \nonumber\\
    &\mapsto
    \sum_{j=0}^{N-1}
    g(x_{\alpha,j})
    \frac{I_{\alpha,j}-Z_{\alpha,j}}{2}.
\end{align}
Therefore, the one-hot encoded potential term becomes,
\begin{equation}
    D_{g,\alpha}
    =
    \sum_{j=0}^{N-1}
    g(x_{\alpha,j})
    \frac{I_{\alpha,j}-Z_{\alpha,j}}{2}.
\end{equation}
Thus, a one-variable diagonal function is translated into a weighted sum of single-qubit occupation operators.

More generally, suppose the objective function is decomposed into grouped terms,
\begin{equation}
    f(x)=\sum_m f_m(x_{S_m}),
\end{equation}
where $S_m$ is the subset of variables on which $f_m$ has support and $s_m=|S_m|$. For a support-$s_m$ term, each point on the supported grid is selected by the product of the corresponding one-hot occupation projectors. The potential operator becomes
\begin{align}
  & D_{f_m}
  = \sum_{\{j_\alpha:\alpha\in S_m\}} \biggl\{
  f_m\left(\{x_{\alpha,j_\alpha}\}_{\alpha\in S_m}\right) \nonumber \\
  & \times \prod_{\alpha\in S_m}
  \left(
    \frac{I_{\alpha,j_\alpha}-Z_{\alpha,j_\alpha}}{2}
  \right) \biggl\}.
  \label{eq:onehot_supported_potential}
\end{align}
Expanding the product expresses this supported potential term as a sum of Pauli-$Z$ strings with weight at most $s_m$. The corresponding time evolution is then implemented using Pauli-$Z$ rotations.

For a single support-$s_m$ term, the maximum number of non-identity Pauli-$Z$ strings generated by the expansion is,
\begin{equation}
    M_V(s_m) = \sum_{r=1}^{s_m} \binom{s_m}{r}N_r = (N+1)^{s_m}-1 .
    \label{eq:onehot-MV-sm}
\end{equation}
Here, $r$ counts the number of variables included in the Pauli-$Z$ string, while $N_r$ counts the corresponding choices of grid points on those variables.

Before merging identical Pauli strings across different objective terms, the total number of potential rotations per potential layer satisfies,
\begin{equation}
    M_V \le \sum_m \left[ (N+1)^{s_m}-1 \right].
    \label{eq:onehot-MV}
\end{equation}
If $G$ is the number of grouped objective terms and
\begin{equation}
    s_{\max}=\max_m s_m ,
\end{equation}
then
\begin{equation}
    M_V = O(GN^{s_{\max}}).
    \label{eq:onehot-MV-scaling}
\end{equation}
Thus, the potential-synthesis cost is controlled by the largest objective-term support, rather than by the full dimension $d$, unless the objective contains fully coupled $d$-variable terms.

A weight-$w$ Pauli-$Z$ rotation can be implemented independently using a CNOT ladder with approximately $2(w-1)$ CNOT gates and one arbitrary $R_z$ rotation. Since the maximum string weight in the one-hot potential is $s_{\max}$, independent synthesis gives
\begin{equation}
    N_{\mathrm{CX}}^{(V)} = O(N_t s_{\max}M_V).
\end{equation}
When $s_{\max}$ is small, this is effectively linear in the number of potential rotations.

More optimized diagonal-synthesis methods, including parity-network, phase-polynomial, and phase-gadget synthesis, can reduce the CNOT overhead by computing several parities within a shared CNOT network rather than synthesizing each $Z$ string independently~\cite{Amy2017CNOTPhase, Cowtan2019PhaseGadget, Vandaele2021PhasePolynomial}. These methods primarily optimize the Clifford entangling gate layer and may also reduce CNOT depth under hardware-connectivity constraints. However, they do not reduce the number of distinct phase angles required for a generic dense potential. After identical strings are merged and zero coefficients are removed, each remaining nonzero Pauli-$Z$ string still carries an independent coefficient and therefore requires one associated $R_z$ rotation. Therefore, we keep the $O(M_V)$ rotation count unchanged, while the CNOT prefactor can depend on the chosen diagonal-synthesis backend.

\subsection{Total rotation and fault-tolerant \texorpdfstring{$T$}{T}-gate count}
\label{sec:one-hot:t-count}

For first-order Trotterization, one QHD time step has the form,
\begin{equation}
    U_s \approx e^{-i\Delta t H_V(t_s)} e^{-i\Delta t H_K(t_s)} .
\end{equation}
The number of Pauli rotations per time step is therefore,
\begin{equation}
    M_{\mathrm{step}}^{(1)} = M_K+M_V .
\end{equation}
For $N_t$ QHD time steps, the total number of arbitrary rotations is,
\begin{equation}
    N_{\mathrm{rot}}^{(1)} = N_t(M_K+M_V) = O\!\left( N_t[2dN+GN^{s_{\max}}] \right).
    \label{eq:onehot-Nrot-first}
\end{equation}

Similarly, for second-order Trotterization,
\begin{equation}
    U_s \approx e^{-i\frac{\Delta t}{2}H_V(t_s)} e^{-i\Delta t H_K(t_s)} e^{-i\frac{\Delta t}{2}H_V(t_s)} .
\end{equation}
Naively, this gives,
\begin{equation}
    N_{\mathrm{rot}}^{(2)} = N_t(M_K+2M_V).
\end{equation}
However, adjacent potential evolutions are diagonal and commute with each other. The two neighboring half-step potential layers may therefore be merged, reducing the number of distinct potential layers from $2N_t$ to approximately $N_t+1$. Hence,
\begin{equation}
    N_{\mathrm{rot}}^{(2)} \approx N_tM_K+(N_t+1)M_V .
\end{equation}
At the scaling level, both first- and second-order formulas give,
\begin{equation}
    N_{\mathrm{rot}} = O\!\left( N_t[2dN+GN^{s_{\max}}] \right).
    \label{eq:one-hot-nrot}
\end{equation}

In a fault-tolerant Clifford+$T$ implementation, arbitrary rotations dominate the non-Clifford cost because they are not native logical gates and must be approximated by gate synthesis. Let $C_T(\epsilon_{\mathrm{rot}})$ denote the number of $T$ gates required to synthesize one arbitrary rotation to precision $\epsilon_{\mathrm{rot}}$. For a synthesis method such as Gridsynth~\cite{ross2014optimal},
\begin{equation}
    C_T(\epsilon_{\mathrm{rot}}) = 3\log(1/\epsilon_{\mathrm{rot}}) + O\!\left[ \log\log(1/\epsilon_{\mathrm{rot}}) \right].
\end{equation}
Thus, to leading order,
\begin{equation}
    C_T(\epsilon_{\mathrm{rot}}) = O\!\left( \log(1/\epsilon_{\mathrm{rot}}) \right).
    \label{eq:ct_factor}
\end{equation}
The total $T$ count is then approximately,
\begin{equation}
    N_T
    \approx
    N_{\mathrm{rot}} C_T(\epsilon_{\mathrm{rot}}).
\end{equation}

If the total error budget assigned to rotation synthesis is $\epsilon_r$, a simple uniform allocation gives
\begin{equation}
    \epsilon_{\mathrm{rot}} = \frac{\epsilon_r}{N_{\mathrm{rot}}}.
    \label{eq:ep_rot}
\end{equation}
Substituting this into the synthesis cost gives
\begin{equation}
    N_T
    =
    O\!\left(
        N_{\mathrm{rot}}
        \log
        \frac{N_{\mathrm{rot}}}{\epsilon_r}
    \right).
\end{equation}
Using Eq.~\eqref{eq:one-hot-nrot}, the one-hot QHD $T$-gate scaling becomes
\begin{align}
    N_T
    =
    O\! \bigg(
        &N_t[2dN+GN^{s_{\max}}]
        \nonumber\\
        &\times
        \log
        \frac{
            N_t[2dN+GN^{s_{\max}}]
        }{
            \epsilon_r
        }
    \bigg).
    \label{eq:onehot-T-count}
\end{align}

The $T$ count also constrains the required logical error rate of the consumed magic states. Let $p_T$ be the total error budget assigned to faulty $T$ states. If the computation consumes $N_T$ synthesized $T$ gates, a conservative target for the error per consumed $T$ state is,
\begin{equation}
    p_T^{\mathrm{per}} \lesssim \frac{p_T}{N_T}.
\end{equation}
Thus, increasing the one-hot QHD circuit size has two related effects: it increases the number of synthesized rotations, and hence the number of consumed $T$ states, while also tightening the required error rate of each magic state.

\subsection{Summary} \label{sec:one-hot:summary}

In summary, for one-hot encoded QHD the logical qubit count scales as
\begin{equation}
    Q_{\mathrm{logical}}=O(dN),
\end{equation}
the kinetic rotation count per QHD step scales as
\begin{equation}
    M_K=O(dN),
\end{equation}
and the potential rotation count per potential layer scales as
\begin{equation}
    M_V=O(GN^{s_{\max}}),
\end{equation}
where $s_{\max}$ is the maximum support size of the objective-function terms. 
The total number of arbitrary rotations scales as
\begin{equation}
    N_{\mathrm{rot}}
    =
    O\!\left(
        N_t[dN+GN^{s_{\max}}]
    \right),
\end{equation}
and the corresponding $T$ count scales as
\begin{equation}
    N_T
    =
    O\!\left(
        N_{\mathrm{rot}}
        \log
        [N_{\mathrm{rot}}/\epsilon_r]
    \right).
\end{equation}
Consequently, the one-hot encoding has benign kinetic scaling, while the potential cost is governed by the largest variable support appearing in the objective. 
For separable objectives, where $s_{\max}=1$, the total circuit size scales essentially as $O(N_t dN)$. 
For pairwise objectives, where $s_{\max}=2$, the potential cost scales as $O(N_t GN^2)$ and may dominate. 
For fully coupled objectives, the potential cost can scale as $O(N_t N^d)$, reflecting the intrinsic cost of representing a generic dense $d$-dimensional potential on a grid.

\section{QHD Binary amplitude encoding and resource scaling}
\label{sec:binary}

In binary amplitude encoding, each coordinate is represented using
\begin{equation}
    b=\log_2 N
\end{equation}
qubits, where $N$ is the number of grid points per variable.  The grid index is encoded directly in the computational basis,
\begin{equation}
    \ket{j} \mapsto \ket{j_{b-1}\cdots j_1 j_0}, \qquad j\in\{0,\ldots,N-1\}.
\end{equation}
For an optimization problem with $d$ variables, hence the number of logical data qubits is,
\begin{equation}
    n_{\binary} = db = d\log_2 N,
\end{equation}
contracting to the one-hot data-qubit requirement $n_{\onehot}=dN$. Thus, the qubit-reduction factor of binary encoding relative to one-hot encoding is,
\begin{equation}
    \frac{ n_{\onehot} }{ n_{\binary} } = \frac{N}{\log_2 N}.
\end{equation}
The advantage of binary amplitude encoding is therefore its logarithmic qubit scaling in the grid resolution. The cost is that the kinetic and potential operators are no longer represented by simple qubit-local terms.

In the remainder of this section, we analyze the gate-count scaling of binary-encoded QHD. The realization strategies and gate-count scalings for the potential and kinetic terms are presented in Secs.~\ref{sec:binary:potential} and~\ref{sec:binary:kinetic}, respectively. The single-Trotter-step scaling is summarized in Sec.~\ref{sec:binary:one-trotter}, and Sec.~\ref{sec:binary:t-gates} discusses the fault-tolerant $T$-gate scaling. Finally, Sec.~\ref{sec:binary:comparison} compares the binary-encoding realization with the one-hot encoding.

\subsection{Potential operator in binary encoding} \label{sec:binary:potential}

The potential term is diagonal in the computational basis,
\begin{equation}
    H_V(t) = e^{\chi(t)} f(\mathbf{x}).
\end{equation}
As in the one-hot construction, assume that the objective function is decomposed into grouped terms,
\begin{equation}
    f(\mathbf{x}) = \sum_m f_m(x_{S_m}),
\end{equation}
where $S_m$ is the subset of variables appearing in term $m$, and $s_m=|S_m|$ is the variable support size.  A support-$s_m$ term acts on $n_m$ qubits, where
\begin{equation}
    n_m = s_m b = s_m\log_2 N.
\end{equation}
The corresponding diagonal operator has $2^{n_m} = N^{s_m}$ diagonal entries.

For each grouped objective term $f_m$, let $D_m$ denote the diagonal operator acting on the $n_m$ supported binary qubits whose computational-basis diagonal entries are the sampled values of $f_m$,
\begin{equation}
    D_m = \sum_{x\in\{0,1\}^{n_m}} f_m(x_{S_m})\dyad{x_{S_m}}.
\end{equation}
The scalar prefactor $e^{\chi(t)}$ in $H_V(t)$ is applied later to the rotation angles, so it is not included in the definition of $D_m$ for now.

Any such diagonal operator on $n_m$ qubits can be expanded in the Pauli-$Z$ basis. Let $\{n_m\}=\{1,2,\ldots,n_m\}$ denote the set of qubit labels in the supported binary registers. Each subset $A \subseteq \{n_m\}$ specifies one Pauli-$Z$ string, i.e., the product of $Z$ operators on the qubits in $A$ and identities on all other qubits, and hence
\begin{equation}
    D_m = \sum_{A\subseteq \{n_m\}} c_A Z_A, \quad Z_A=\prod_{j\in A} Z_j .
\end{equation}
The empty subset $A=\emptyset$ corresponds to the identity operator. The coefficients $c_A$ are obtained by applying the Walsh-Hadamard transform to the diagonal entries of $D_m$ as
\begin{equation}
    c_A = \frac{1}{2^{n_m}} \sum_{x\in\{0,1\}^{n_m}} f_{m}(x_{S_m}) (-1)^{x\cdot A}.
\end{equation}
The identity term contributes only a global phase and can be dropped. Thus a generic support-$s_m$ potential block contains up to $(N^{s_m}-1)$ nontrivial $Z$-string rotations~\cite{Hadfield2021}.

The number of arbitrary potential rotations per potential layer is then bounded by,
\begin{equation}
    M_V^{\mathrm{bin}} \le \sum_m \left( N^{s_m}-1 \right).
\end{equation}
Similar to the discussion in Sec.~\ref{sec:one-hot:potential-count}, by defining $s_{\max}=\max_m s_m$ and $G$ is the number of grouped objective terms, the binary rotation gate count scaling from the potential term can be estimated as,
\begin{equation}
    M_V^{\mathrm{bin}} = O(GN^{s_{\max}}).
\end{equation}

Each nonzero Pauli-$Z$ coefficient generates a rotation
\begin{equation}
    \exp\left[ -i\Delta t\, e^{\chi(t)} c_A Z_A \right] = R_{Z_A}(\theta_A),
\end{equation}
with
\begin{equation}
    \theta_A = 2\Delta t\,e^{\chi(t)}c_A,
\end{equation}
using the convention $R_P(\theta) = \exp\left(-i \theta P / 2\right)$.

The CNOT cost of synthesizing these rotations depends on the diagonal synthesis methods.  If every $Z$-string rotation is compiled independently, a weight-$w$ string requires approximately $2(w-1)$ CNOT gates and one arbitrary $R_z$ rotation.  Since a support-$s_m$ binary potential term can have weight as large as
\begin{equation}
    w_{\max} = s_m\log_2 N.
\end{equation}
Independent synthezing each of the Pauli-rotation can give
\begin{equation}
    N_{\mathrm{CNOT}}^{V,\mathrm{bin,direct}} = O\!\left(
        s_{\max}\log(N)\,G N^{s_{\max}}
    \right).
\end{equation}
Using a parity-network or phase-polynomial synthesis method, the CNOT network can be shared across many diagonal rotations. For a dense diagonal unitary matrix with dimension $N$, the CNOT cost can be $O(N)$~\cite{Welch_2014, Zhuang2024Depth}, the total CNOT cost for the potential term can be modeled as,
\begin{equation}
    N_{\mathrm{CNOT}}^{V,\mathrm{bin,parity}} = O(GN^{s_{\max}})
\end{equation}
up to constant factors depending on the synthesizing method. However, the arbitrary rotation count remains
\begin{equation}
    N_{\mathrm{rot}}^{V,\mathrm{bin}} = O(GN^{s_{\max}}).
    \label{eq:binary_potential_rot}
\end{equation}

\subsection{Kinetic operator using the exact QFT construction} \label{sec:binary:kinetic}

The binary kinetic operator is treated differently from the one-hot kinetic operator.  In one-hot encoding, the finite-difference hopping terms map directly to local nearest-neighbor $XX+YY$ interactions. In binary amplitude encoding, the transition $\dyad{j}{j+1}$ connects two $b$-bit computational basis states and is not a local two-qubit hopping term anymore.

Other constructions of the finite-difference Laplacian are also possible. For example, one may synthesize the sparse transition structure of the Laplacian directly, use binary-reflected Gray-code encoding to reduce the gate cost for realizing the kinetic terms~\cite{Chang2022GrayCode}, or use circuit constructions developed for PDE simulation and Schr\"odingerisation-based algorithms~\cite{Hu2024Schrodingerisation}. One may also replace the exact QFT with an approximate QFT (AQFT) to reduce the number of controlled rotations. These alternatives change the detailed gate counts and can potentially provide a more resource-favorable QHD implementation. In the present subsection, however, we focus on the exact QFT-based kinetic implementation. Alternative kinetic term realizations with approximations are discussed separately in later subsections.

For periodic boundary conditions, the finite-difference Laplacian can instead be diagonalized exactly by the quantum Fourier transform.  For one variable, define
\begin{equation}
    U_{\mathrm{QFT}}\ket{j} = \frac{1}{\sqrt{N}} \sum_{k=0}^{N-1} e^{2\pi i jk/N}\ket{k}.
\end{equation}
The periodic finite-difference Laplacian is diagonal in the Fourier
basis,
\begin{equation}
    D_L \equiv U_{\mathrm{QFT}} L_h U_{\mathrm{QFT}}^\dagger = \sum_{k=0}^{N-1} \lambda_k \dyad{k},
\end{equation}
where the eigenvalues $\lambda_k$ is,
\begin{equation}
    \lambda_k = -\frac{4}{h^2} \sin^2\left(\frac{\pi k}{N}\right).
    \label{eq:binary_laplacian_eigenvalues}
\end{equation}
For Dirichlet boundary conditions, the Laplacian can instead be diagonalized using the quantum discrete sine transform~\cite{klappenecker2001QDST}.

For the QHD kinetic Hamiltonian, the single-register kinetic evolution can be written as
\begin{equation}
    U_K^{(1)}(t_s) = U_{\mathrm{QFT}}^\dagger \exp\left[
        i\frac{e^{\phi(t_s)}\Delta t}{2} D_L
    \right]
    U_{\mathrm{QFT}}.
\end{equation}
For $d$ variables, the kinetic Hamiltonian is a sum of commuting single-coordinate Laplacians.  Thus, the kinetic layer can be implemented by applying the QFT to each coordinate register, applying the single-register Fourier-basis phase function to each register, and then applying the inverse QFT to each coordinate register, i.e.,
\begin{equation}
    U_K(t_s) = \prod_{\alpha=1}^d \left[
        U_{\mathrm{QFT},\alpha}^\dagger
        \exp\left(
            i\frac{e^{\phi(t_s)}\Delta t}{2} D_{L,\alpha}
        \right)
        U_{\mathrm{QFT},\alpha}
    \right]. \nonumber
\end{equation}
The factors commute because each acts on a different coordinate
register.

\subsubsection{Exact QFT cost}

The exact QFT on $b$ qubits contains
\begin{equation}
    N_{\mathrm{CR}}^{\mathrm{QFT}}(b) = \frac{b(b-1)}{2}
\end{equation}
controlled-phase rotations, together with $b$ Hadamard gates and, depending on convention, $\lfloor b/2\rfloor$ final bit-reversal swaps. If the bit reversal is handled by classical relabeling or absorbed into subsequent circuit layout, these swaps need not be physically implemented~\cite{nielsen2010quantum}.

Using a standard controlled-$R_z$ decomposition, each controlled rotation requires two CNOT gates and two arbitrary single-qubit $R_z$ rotations. Thus, one QFT contributes
\begin{align}
    N_{\mathrm{CNOT}}^{\mathrm{QFT}}(b) &= 2N_{\mathrm{CR}}^{\mathrm{QFT}}(b) = b(b-1), \\ 
    N_{\mathrm{rot}}^{\mathrm{QFT}}(b) &= 2N_{\mathrm{CR}}^{\mathrm{QFT}}(b) = b(b-1).
\end{align}
If final bit-reversal swaps are implemented explicitly, they add $3\lfloor b/2\rfloor$ CNOT gates per QFT. Equivalently, writing the QFT in terms of controlled-phase gates
\begin{equation}
    \operatorname{CP}(\theta) = \operatorname{diag}(1,1,1,e^{i\theta})
\end{equation}
may introduce a different constant number of one-qubit $R_z$ rotations per controlled phase. These convention-dependent choices do not change the $O(b^2)$ scaling.

A kinetic layer uses one QFT and one inverse QFT per coordinate register. Therefore, the QFT and inverse-QFT contributions to the kinetic CNOT count are
\begin{equation}
    N_{\mathrm{CNOT}}^{K,\mathrm{QFT}} = 2d\,N_{\mathrm{CNOT}}^{\mathrm{QFT}}(b)
    = O(db^2) = O\!\left(d\log^2 N\right),
\end{equation}
and the corresponding arbitrary rotation count is
\begin{equation}
    N_{\mathrm{rot}}^{K,\mathrm{QFT}} = 2d\,N_{\mathrm{rot}}^{\mathrm{QFT}}(b)
    = O(db^2) = O\!\left(d\log^2 N\right).
    \label{eq:binary_qft_rot}
\end{equation}
Hadamard gates, CNOT gates, and swaps are Clifford-level operations and are not included in the arbitrary rotation count.

\subsubsection{Fourier-basis kinetic phase cost}

The diagonal Fourier-basis kinetic phase for one coordinate is
\begin{equation}
    \exp\left[
        i\frac{e^{\phi(t_s)}\Delta t}{2}
        D_L
    \right] = \sum_{k=0}^{N-1}
    \exp\left[
        i\frac{e^{\phi(t_s)}\Delta t}{2}
        \lambda_k
    \right] \dyad{k}. \nonumber
\end{equation}
This is a diagonal operator on $b=\log_2 N$ qubits whose diagonal entries are determined by $\lambda_k$ given in Eq.~\eqref{eq:binary_laplacian_eigenvalues}. Similar to the potential term constructions, this phase function can be expanded into Pauli-$Z$ strings. 
After dropping the identity contribution, the number of nontrivial diagonal rotations is at most $2^b-1 = N-1$. Thus, the Fourier-basis kinetic phase contributes
\begin{equation}
    N_{\mathrm{rot}}^{K,\mathrm{phase}} = O(dN)
\end{equation}
arbitrary rotations per kinetic layer.

The CNOT cost of the Fourier-basis phase depends on the diagonal synthesis backend, similar to the potential term realization. Independent synthesis of all $Z_A$-string rotations gives
\begin{equation}
    N_{\mathrm{CNOT}}^{K,\mathrm{phase,direct}} = O(dN\log N),
\end{equation}
because the maximum string weight is $b=\log_2 N$.  With a phase-polynomial or parity-network implementation for the dense single-register diagonal~\cite{Amy2017CNOTPhase, Vandaele2021PhasePolynomial}, the CNOT count can be further reduced by sharing the Clifford parity-computation circuits across different Pauli rotations,
\begin{equation}
    N_{\mathrm{CNOT}}^{K,\mathrm{phase,parity}} = O(dN),
\end{equation}
again up to backend-dependent constants.

Combining the exact-QFT layers with the Fourier-basis diagonal phase, the kinetic arbitrary rotation count per kinetic layer is
\begin{align}
    N_{\mathrm{rot}}^{K,\mathrm{bin}} & = N_{\mathrm{rot}}^{K,\mathrm{QFT}} + N_{\mathrm{rot}}^{K,\mathrm{phase}} \nonumber \\
    & = O\!\left( d\log^2 N+dN \right).
\end{align}
For large number of grid points, $N$ can be typically larger than $\log^2 N$, the leading order becomes
\begin{equation}
    N_{\mathrm{rot}}^{K,\mathrm{bin}} = O(dN),
    \label{eq:binary_kin_rot}
\end{equation}
but we will explicitly keep the exact-QFT contribution when comparing against AQFT or alternative kinetic constructions.

The corresponding CNOT count is
\begin{equation}
    N_{\mathrm{CNOT}}^{K,\mathrm{bin,direct}} = O\!\left( d\log^2 N+dN\log N \right)
\end{equation}
for independent synthesis of the Fourier-basis phase rotations, and
\begin{equation}
    N_{\mathrm{CNOT}}^{K,\mathrm{bin,parity}} = O\!\left( d\log^2 N+dN \right)
\end{equation}
for a parity-network diagonal synthesis backend.

\subsection{Resource scaling for a single time step} \label{sec:binary:one-trotter}

As the rotation gates are the dominant cost in a FT implementation after rotation synthesis, we first summarize the arbitrary-rotation counts. The CNOT gate counts depend more strongly on the chosen diagonal-synthesis backend.

For the first-order Trotter approximation,
\begin{equation}
    U_s \approx e^{-i H_V(t_s) \dt} e^{-i H_K(t_s) \dt},
\end{equation}
the rotation count is obtained by adding the potential contribution and the two kinetic contributions from the exact QFTs and the Fourier-basis phase
\begin{equation}
    N_{\mathrm{rot,step}}^{\mathrm{bin}} = N_{\mathrm{rot}}^{V,\mathrm{bin}} + N_{\mathrm{rot}}^{K,\mathrm{QFT}} + N_{\mathrm{rot}}^{K,\mathrm{phase}}.
\end{equation}
Using the estimates in the previous subsections [see Eqs.~\eqref{eq:binary_potential_rot},~\eqref{eq:binary_qft_rot}, and~\eqref{eq:binary_kin_rot}],
\begin{equation}
    N_{\mathrm{rot,step}}^{\mathrm{bin}} = O\!\left( GN^{s_{\max}} + d\log^2 N + dN \right).
\end{equation}
For $N_t$ time steps, the total rotation gates become 
\begin{equation}
    N_{\mathrm{rot}}^{\mathrm{bin}} = O\!\left( N_t \left[
            GN^{s_{\max}} + d\log^2 N + dN
        \right] \right).
        \label{eq:nrot_bin_first_trotter}
\end{equation}

For a second-order product formula,
\begin{equation}
    U_s
    \approx
    e^{-i\frac{\Delta t}{2}H_V(t_s)}
    e^{-i\Delta t H_K(t_s)}
    e^{-i\frac{\Delta t}{2}H_V(t_s)},
\end{equation}
the kinetic contribution is unchanged. Similar to the one-hot construction, adjacent potential half-layers are still diagonal and commute with each other, so they can be merged across step boundaries. Thus the potential contribution contains $N_t+1$ effective layers rather than $2N_t$ half-layers. Therefore, the total number of rotation gates is
\begin{equation}
    N_{\mathrm{rot}}^{\mathrm{bin}} = O\!\left(
        (N_t+1)GN^{s_{\max}} + N_t\left[d\log^2 N+dN\right]
    \right).
    \label{eq:nrot_bin_second_trotter}
\end{equation}

The corresponding CNOT estimates can be obtained by replacing each rotation block above with the appropriate CNOT gate-count model. For direct independent synthesis of all Pauli-$Z$ rotations, the leading total CNOT scaling becomes,
\begin{align}
    N_{\mathrm{CNOT}}^{\mathrm{bin}} =
    & O\!\left( N_t\left[ Gs_{\max}\log(N)\,N^{s_{\max}} \right. \right.
    \nonumber \\
    & \qquad \ \left. \left. + d\log^2 N + dN\log N \right] \right) 
\end{align}
for first-order Trotterization, with the replacement $N_tG\mapsto (N_t+1)G$ in the potential term for merged second-order half-layers.  With parity-network synthesis for the dense diagonal blocks, the analogous scaling is
\begin{align}
    N_{\mathrm{CNOT}}^{\mathrm{bin,parity}} = & O\!\left( N_t\left[ GN^{s_{\max}} + d\log^2 N 
    + dN \right] \right),
\end{align}
again with $N_t G \rightarrow (N_t+1) G$ for merged second-order potential time steps.

\subsection{Fault-tolerant rotation synthesis and T-count model} \label{sec:binary:t-gates}

The $T$-count is dominated by the synthesis of arbitrary rotations, together with any non-Clifford overhead used by the chosen diagonal synthesis and QFT decomposition. We again use $C_T(\epsilon_{\mathrm{rot}})$ denote the $T$-count required to synthesize one arbitrary single-qubit rotation to precision $\epsilon_{\mathrm{rot}}$. As we discussed in Sec.~\ref{sec:one-hot:t-count}, the scaling for the pre-factor $C_T$ can be estimated by Eq.~\eqref{eq:ct_factor}.

For the exact-QFT kinetic implementation, arbitrary rotations arise from two sources, (1) the controlled-phase gates in the exact QFT and inverse QFT, (2) the diagonal Fourier-basis kinetic phase $\exp(i e^{\phi(t_s)}\Delta t D_L/2)$. The potential term contributes arbitrary $Z$-string rotations from the Walsh-Hadamard expansion of each diagonal potential block.

If the total synthesis error budget is again $\epsilon_r$, and this budget is allocated uniformly over all arbitrary rotations, then the synthesis error per rotation gate can be similarly estimated by Eq.~\eqref{eq:ep_rot}, where the total rotation gate count is given by Eqs.~\eqref{eq:nrot_bin_first_trotter} for first-order trotter approximation and~\eqref{eq:nrot_bin_second_trotter} for second order. Therefore the number of $T$-count becomes,
\begin{equation}
    N_T^{\mathrm{bin}} = O\!\left( N_{\mathrm{rot}}^{\mathrm{bin}} \log \frac{N_{\mathrm{rot}}^{\mathrm{bin}} }{\epsilon_{\mathrm{synth}}^{\mathrm{tot}} }\right),
\end{equation}
where the rotation gate count $N_{\mathrm{rot}}^{\mathrm{bin}}$ are given in Eqs.~\eqref{eq:nrot_bin_first_trotter} and~\eqref{eq:nrot_bin_second_trotter}.

\begin{table*}[t]
\caption{ Asymptotic gate-count estimates for one Trotter step of the kinetic and potential operators under one-hot and binary encodings. Here $d$ is the number of degrees of freedom, $N$ is the number of grid points per degree of freedom, $G$ is the number of potential-function terms, and $s_{\max}$ is the maximum support size of a potential term. For the binary encoding methods with parity network (QFT-PN) realization and using approximated QFT and using $k^2$ to approximate the eigenvalues in the momentum space (QFT-$k^2$), ``--'' means the scaling will be identical to the QFT-exact if no other optimizations are applied.}
\label{tab:encoding_gate_scaling}
\begin{ruledtabular}
\begin{tabular}{lcccc}
Encoding
&
\multicolumn{2}{c}{Kinetic operator}
&
\multicolumn{2}{c}{Potential operator}
\\
\cline{2-3}
\cline{4-5}
&
$R_z$ gates
&
CX gates
&
$R_z$ gates
&
CX gates
\\
\hline
One-hot
&
$\mathcal{O}(dN)$
&
$\mathcal{O}(dN)$
&
$\mathcal{O}\!\left(G N^{s_{\max}}\right)$
&
$\mathcal{O}\!\left(s_{\max} G N^{s_{\max}}\right)$
\\
Binary, QFT-exact
&
$\mathcal{O}\!\left(d\log^2 N + dN\right)$
&
$\mathcal{O}\!\left(dN\log N + d\log^2 N\right)$
&
$\mathcal{O}\!\left(G N^{s_{\max}}\right)$
&
$\mathcal{O}\!\left(s_{\max}\log N \, G N^{s_{\max}}\right)$
\\
\hline
Binary, QFT-PN
&
--
&
$\mathcal{O}\!\left(dN+d\log^2 N\right)$
&
--
&
$\mathcal{O}\!\left(GN^{s_{\max}}\right)$
\\
Binary, AQFT-$k^2$
&
$\mathcal{O}\!\left(d\log^2 N + d \log N\right)$
&
$\mathcal{O}\!\left(d\log^2 N+d\log N\right)$
&
--
&
--
\\
\end{tabular}
\end{ruledtabular}
\end{table*}

\subsection{Cost comparison to one-hot encoding} \label{sec:binary:comparison}

We now compare the leading gate-count scalings of the binary and one-hot encodings under the implementations considered above. With our current construction, in both cases, the FT $T$ count is dominated by the synthesis of arbitrary rotation gates rather than by Toffoli-heavy arithmetic or other oracle subroutines. It is therefore most useful to compare the arbitrary-rotation counts directly. Since the first- and second-order product formulas have the same leading scaling up to constant factors and only need to increase the time step $N_t$ by $(N_t+1)$ in the potential layers, we focus on the first-order estimates.

The gate scalings for a single Trotter step of both the one-hot encoding and the binary finite-difference Laplacian realizations are summarized in Table~\ref{tab:encoding_gate_scaling}. For one-hot encoding, the first-order rotation count in Eq.~\eqref{eq:one-hot-nrot} consists of a nearest-neighbor kinetic contribution and a potential contribution. For binary encoding with the exact-QFT kinetic realization, the leading-order rotation gates arise from synthesizing the potential terms and kinetic phases, as well as from the QFT and inverse-QFT operations. At the level of arbitrary rotations, the two encodings have the same leading dependence on the grid size for large $N$. Specifically, realizing the potential terms requires the same scaling of rotation gates, while for the kinetic operators, the leading-order contribution of the binary realization comes from the kinetic-phase realization.

The $O(d N)$ kinetic-phase rotation count in the exact-QFT construction comes from treating the finite-difference eigenvalues in Eq.~\eqref{eq:binary_laplacian_eigenvalues} as a generic diagonal function of the Fourier index $k$ (wave number in momentum space). This generic diagonal representation does not exploit additional algebraic structure, so in the worst case, one keeps up to $N-1$ nontrivial Pauli-$Z$ components per coordinate register. A physically motivated simplification is to focus on the low-momentum, long-wavelength sector of the kinetic spectrum. In QHD, the initial state is often chosen from the ground state or a low-energy state of an initial Hamiltonian with a kinetic term, and for a light-particle kinetic scale, this state is expected to have most of its support on low-energy momentum modes. These modes are also the smooth components of the wavefunction and can be the most relevant part of the dynamics when the evolution is intended to remain close to the low-energy sector. In this regime, the finite-difference dispersion can be approximated by the continuum free-particle form,
\begin{equation}
    \lambda_k = - \frac{4}{h^2}\sin^2\!\left(\frac{\pi k}{N}\right) \approx -\frac{4\pi^2}{h^2N^2} k^2 .
    \label{eq:k2}
\end{equation}

Under this approximation, the Fourier-basis kinetic phase is no longer a generic diagonal function. Writing the binary expansion of the momentum index as
\begin{equation}
    k=\sum_{\ell=0}^{b-1}2^\ell k_\ell, \quad b=\log_2N,
\end{equation}
where $k_\ell$ is $\ell$-th bit of the number $k$ written in the binary format. With this binary representation, the quadratic function contains only one-bit and two-bit terms,
\begin{equation}
    k^2 = \sum_{\ell=0}^{b-1}2^{2\ell}k_\ell + 2\sum_{0\leq \ell<m<b}2^{\ell+m}k_\ell k_m .
\end{equation}
To realize these terms on the binary-encoded state, a single-bit term $k_\ell$ corresponds to the projector $\dyad{1_\ell}$, which can be written as a single Pauli-$Z$ term up to an identity contribution. Similarly, a two-bit term $k_\ell k_m$ corresponds to the projector $\dyad{1_\ell,1_m}$, which expands into Pauli strings with at most a $Z_\ell Z_m$ component. Consequently, the kinetic phase can be synthesized using only single-qubit $Z$ rotations and two-qubit $ZZ$ rotations, rather than a dense set of higher-weight Pauli-$Z$ strings. The number of such phase rotations is therefore $O(b^2)=O(\log^2 N)$ per coordinate register, giving $O(d\log^2 N)$ kinetic-phase rotations for $d$ variables.

In addition to the low-momentum approximation, if the surrounding QFT and inverse-QFT are implemented using an AQFT subroutine with $O(b)$ retained controlled rotations per register, their additional rotation and CNOT costs scale as $O(d \log N)$~\cite{Nam2020AQFT}. Thus, an approximate-QFT plus $k^2$ kinetic-phase backend can reduce the binary kinetic scaling from the generic exact-QFT estimate $O(dN+d\log^2N)$ rotations to $O(d\log^2N+d\log N)$ rotations, with the same asymptotic scaling for the associated CNOT count up to constant factors from the chosen $ZZ$-rotation decomposition. The gate count scaling for the AQFT with $k^2$ approximated Laplacian operator realization is also summarized in Table~\ref{tab:encoding_gate_scaling}.

The comparison above shows that the baseline one-hot and exact-QFT binary implementations have similar leading arbitrary-rotation scaling, while binary encoding uses only $O(d\log N)$ data qubits compared with $O(dN)$ for one-hot encoding at the same grid resolution. The tradeoff is that the binary diagonal operators are less local in the qubit basis. Without parity-network or phase-polynomial optimization, independently compiling each binary Pauli-$Z$ rotation introduces an additional string-weight factor in the CNOT count. This is the origin of the $s_{\max}\log N$ factor in the direct binary potential estimate. In addition, when the connectivity of the quantum hardware is limited, the binary realization can cause more CNOT gates for routing in compilation. In contrast, one-hot potential strings have weight at most $s_{\max}$. Therefore, at the level of unoptimized CNOT counts, one-hot encoding can look more favorable, even though its logical data-register size is much larger.

In a fault-tolerant setting based on Clifford+$T$ synthesis, however, the CNOT count is usually a less direct bottleneck than the number of non-Clifford gates. As we discussed briefly in Sec.~\ref{sec:one-hot:encoding}, logical Clifford operations, including CNOTs, are typically much cheaper than synthesized arbitrary rotations or distilled $T$ gates. Under this cost model, binary encoding is attractive because it gives comparable leading rotation and $T$-count scaling while reducing the data-register size from $O(dN)$ to $O(d\log N)$. Moreover, the AQFT-$k^2$ kinetic operator realization method discussed above can potentially reduce the binary kinetic rotation and CNOT scaling from the generic exact-QFT estimate to polylogarithmic scaling in $N$. This provides an additional route by which binary encoding may outperform the one-hot strategy, provided that the low-momentum approximation is accurate for the relevant low-momentum sector and that connectivity overheads do not dominate. We will test the accuracy and practical impact of this AQFT-$k^2$ kinetic approximation numerically in later sections.

The above comparison assumes a fault-tolerant architecture in which logical qubits can be arranged in a sufficiently uniform local layout, as in an idealized surface-code implementation where high-weight Pauli rotations can be supported by multi-patch lattice-surgery operations~\cite{Litinski2019}. In less uniform layouts, the locality of the Pauli rotations may become an important additional cost. Binary potential rotations can involve Pauli-$Z$ strings of weight as large as $s_{\max}\log_2N$, whereas the corresponding one-hot potential strings have weight at most $s_{\max}$. If the logical qubits are distributed across separate chips or modules, or if restricted connectivity makes long parity computations expensive, these binary rotations may require additional routing, lattice-surgery steps, remote gates, or inter-module Pauli measurements. These connectivity-dependent overheads are not included in the asymptotic estimates above and could shift the practical preference toward encodings with lower-weight Pauli rotations in a specific hardware layout.

\section{Quantum Circuit realization and verification} \label{sec:circuits}

In this section, we describe the circuit constructions used in the numerical comparisons. The purpose is to specify how the abstract one-hot and binary QHD Hamiltonians discussed above are transpiled into quantum circuits, which will then provide numerical gate-count complexity analysis.
Specifically, Sec.~\ref{sec:numerics:circuit_construction} discusses how the QHD circuits are constructed and transpiled using one-hot and binary encodings. The gate-based implementation of the exact QHD dynamics is numerically verified on benchmark problems in Sec.~\ref{sec:numerics:exact}, while the implementations using AQFT and low-momentum approximations are verified in Sec.~\ref{sec:numerics:approx}.

\subsection{Circuit Construction}
\label{sec:numerics:circuit_construction}

For both encodings, the circuit construction begins from the same discretized optimization problem defined by the objective function, variable bounds, the resolution (the number of grid points) per coordinate, and a QHD schedule. In the benchmark cases, we focus mainly on optimization target functions with one to two optimization parameters, due to the classical simulation capabilities. We restrict all optimization variables to $(0, 1)$. By default, the resulting kinetic and potential blocks are assembled at midpoint times of the QHD schedule and ordered into either a first-order or second-order Trotter product, consistent with the resource models used in the preceding sections. The transpilation of each potential and kinetic Trotter steps will depend on the encoding methods.

\begin{figure}[t]
    \centering
    \includegraphics[width=0.95\linewidth]{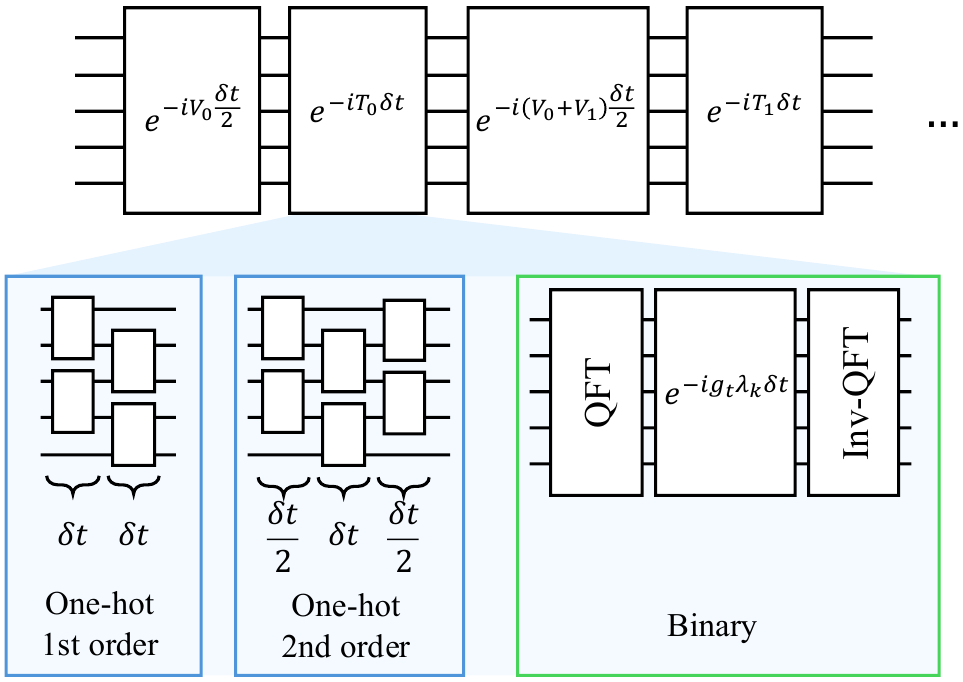}
    \caption{Schematic quantum-circuit structure for the QHD evolution used in the circuit construction. Here $V_i=e^{\chi(t_i)}V$, where $V$ encodes the target objective, and $T_i=e^{\phi(t_i)}K$, where $K$ is the kinetic operator. In both encodings, the potential term is diagonal and can be compiled into Pauli-$Z$ rotations. For one-hot encoding, the kinetic term is mapped to nearest-neighbor $XX+YY$ rotations. We group these Pauli terms into even- and odd-bond layers and apply first-, second-, or fourth-order Trotter decompositions; the first- and second-order realizations are shown. For binary encoding, the QFT and inverse QFT transform the kinetic operator into a diagonal phase operator.}
    \label{fig:qhd_circs}
\end{figure}

\subsubsection{One-hot realization}

In the one-hot construction, each coordinate register contains one qubit per grid point, as in Eq.~\eqref{eq:one-hot-nq}. The circuit acts on the full qubit Hilbert space, but the intended physical subspace is the tensor product of Hamming-weight-one subspaces, one for each coordinate. 

The potential circuit follows directly from the projector representation in Eq.~\eqref{eq:onehot_supported_potential}. For each support-local objective term, the sampled value at a grid configuration is attached to the product of one-hot occupation projectors selecting that configuration. Expanding these projectors gives only Pauli-$Z$ strings, with the identity contribution removed as a global phase. Thus the one-hot potential block is diagonal in the computational basis and is synthesized as a collection of $Z$-string rotations. Repeated strings may be merged before synthesis, and rotations below a prescribed numerical threshold are discarded.

The one-hot kinetic circuit is built from the nearest-neighbor hopping representation summarized in Eq.~\eqref{eq:onehot_even_odd_kinetic}. For each coordinate register, the off-diagonal finite-difference hopping terms are compiled into $XX+YY$ rotations on adjacent grid qubits. Periodic boundary conditions include the wraparound bond, whereas Dirichlet boundary conditions use only the path-graph bonds inside the interval. The diagonal part of the finite-difference Laplacian contributes only an overall phase within the one-hot sector and is therefore omitted from the circuit body using the same global-phase convention described above.

The nearest-neighbor $XX+YY$ terms in the kinetic Hamiltonian do not all commute. Therefore, decomposing the kinetic evolution into sequential two-qubit rotations with the same time step assigned to every bond introduces an additional Trotter error. In the numerical simulations, this error appears as a reduction in the fidelity between the circuit evolution and the reference Schr\"odinger-equation solver. To reduce this error, we partition the $XX+YY$ hopping terms into even- and odd-bond layers and apply first-, second-, or fourth-order Trotter decompositions to the kinetic step. This layered construction is illustrated in Fig.~\ref{fig:qhd_circs}. However, in our benchmarking cases with the selected QHD schedule, second-order Trotterization provides accurate enough simulation outcomes already.

\begin{figure*}[tbp]
    \centering
    \includegraphics[width=0.8 \linewidth]{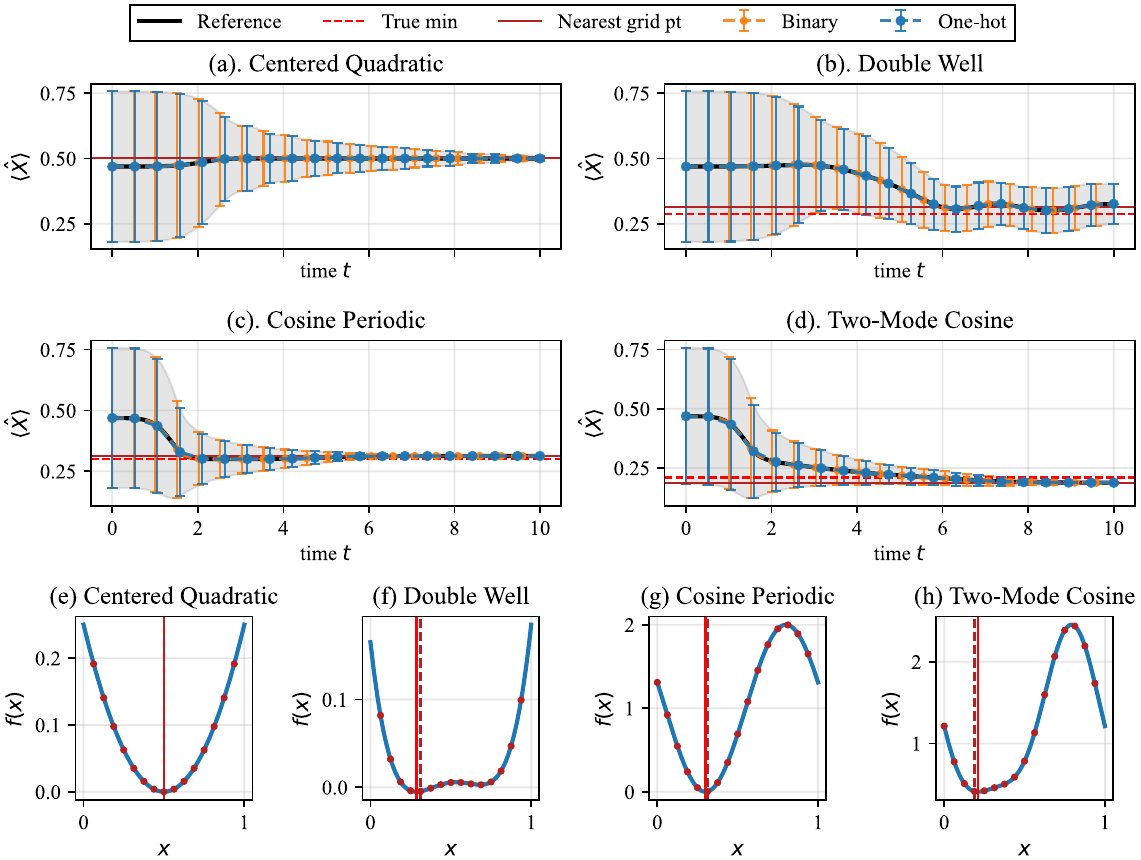}
    \caption{Comparison of the QHD dynamics obtained from the classical split-operator solver and from the one-hot and binary quantum-circuit realizations. The four benchmark target functions $f(x)$ are shown in (e) to (h), and the corresponding position observables are shown in (a) to (d). In panels (a) to (d), the black solid curves show the mean position $\langle x\rangle$, while the black shaded bands indicate $\langle x\rangle\pm\sigma_x$ from the classical Schr\"odinger solver. The blue markers and error bars show the binary-encoding circuit results, and the orange markers and error bars show the one-hot-encoding circuit results, with the error bars indicating $\pm\sigma_x$ extracted from the final simulated state. The dashed red lines indicate the location of the global minimum, while the dark-red solid lines indicate the optimal point on the discretized grid. In (e) to (h), the red dots mark the grid points. We use periodic boundary conditions for all four benchmark problems. }
    \label{fig:verify_circ}
\end{figure*}

\subsubsection{Binary realization}

The binary construction uses $b=\log_2 N$ qubits per coordinate register. A grid point is encoded directly as a computational-basis bit string, so no one-hot subspace constraint is required. The potential term remains diagonal and is compiled by the Pauli-$Z$ expansion described in Sec.~\ref{sec:binary}. In particular, each support-local diagonal block is sampled on its binary grid and transformed into $Z$-string coefficients. The resulting number of nontrivial rotations is given in Eq.~\eqref{eq:binary_potential_rot}.

The kinetic construction differs more substantially between the two encodings. In the baseline binary circuit, each coordinate register is transformed to the Fourier basis, the finite-difference kinetic phase is applied using the eigenvalues in Eq.~\eqref{eq:binary_laplacian_eigenvalues}, and the inverse Fourier transform returns the state to the position basis. The controlled phases in the QFT layers and the Fourier-basis diagonal phase are both treated as arbitrary rotations for the purpose of FT cost estimation, as reflected in Eqs.~\eqref{eq:binary_qft_rot} and~\eqref{eq:binary_kin_rot}. 

For periodic boundary conditions, the finite-difference Laplacian is diagonalized by the standard QFT. For Dirichlet boundary conditions, the Laplacian is instead diagonalized by a discrete sine transform. In this case, we replace the QFT and inverse QFT in the kinetic term by the corresponding quantum sine-transform circuits, implemented using the real-transform routines provided by QRTlib~\cite{ahmadkhaniha2025qrtliblibraryfastquantum}. The kinetic phase angles are changed consistently with the boundary condition. For the interior grids, the periodic eigenvalues in Eq.~\eqref{eq:binary_laplacian_eigenvalues} are replaced by $\lambda_k^{\mathrm{D}}=4h^{-2}\sin^2[\pi(k+1)/(2(N+1))]$, with $k=0,\ldots,N-1$.

In addition, the binary implementation supports an AQFT option and the low-momentum kinetic approximation. In the AQFT routine, a user-specified approximation level sets a cutoff below which controlled-rotation angles are omitted from the QFT circuit. 
The implementation also supports the low-momentum approximation to the kinetic eigenvalues in the Fourier basis.
When the low-momentum approximation is used, the same circuit structure of QFT and inverse QFT parts is retained, but the Fourier-basis kinetic phase is replaced by the quadratic $k^2$ phase discussed in Eq.~\eqref{eq:k2}, reducing the kinetic phase synthesis to one- and two-qubit $Z$ rotations up to the approximation error.

\begin{figure*}[htbp!]
    \centering
    \includegraphics[width=0.85 \linewidth]{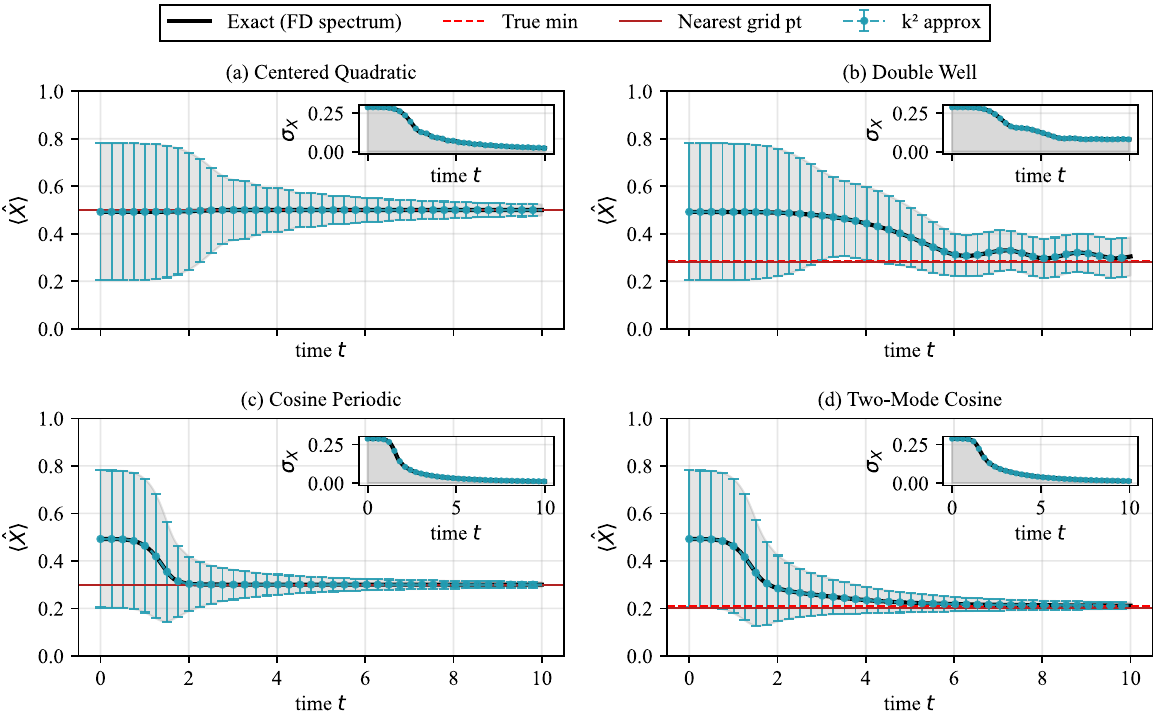}
    \caption{Comparison of the $k^2$-approximated kinetic operator with the exact classical Schr\"odinger-equation solver. We focus on binary encoding for the same four one-dimensional target optimization problems shown in Fig.~\ref{fig:verify_circ}, using $64$ grid points for the QHD simulations. The main panels show the time dependence of $\langle x\rangle$ and the fluctuation $\sigma_x$ compared with the classical solver results, shown as black curves with gray shaded regions. The $k^2$-approximated circuit results are shown as turquoise markers with error bars. The inset in each panel compares $\sigma_x$ from the approximate quantum-circuit realization with that from the exact Schr\"odinger-equation solver. In all four test cases, the approximate circuit implementation agrees well with the exact classical dynamics.}
    \label{fig:verify_k2}
\end{figure*}

\subsection{Circuits for exact QHD implementation verification} \label{sec:numerics:exact}

In this subsection, we verify the generated quantum circuits by comparing their QHD dynamics against a classical Schr\"odinger-equation solver. Specifically, we compare the circuit outputs with a classical split-operator Schr\"odinger-equation solver applied to the same discretized Hamiltonian. The agreement between these two simulations confirms that the circuit construction implements the desired kinetic and potential evolutions and that the resulting dynamics can drive the wavefunction toward the optimal solution of the target objective.

We test the QHD realization with four one-dimensional target problems, shown in Fig.~\ref{fig:verify_circ}e to~\ref{fig:verify_circ}h, to keep the circuit simulations classically tractable. The explicit definitions of the target functions are given in Appendix~\ref{app:benchmark_targets}. Since a one-hot discretization with $N$ grid points requires $N$ qubits for a single variable, we restrict the one-hot verification runs to $N=8$ and $N=16$ for periodic boundary conditions, corresponding to 8- and 16-qubit QHD implementations. For each benchmark and grid resolution, we also run the corresponding binary-encoding circuit to compare the encoded dynamics under the same discretized problem settings.

To drive the QHD dynamics toward low-potential regions, we use a relaxation schedule based on Refs.~\cite{leng2023quantum, wu2026ALQHD},
\begin{equation}
    e^{\phi(t)} = \frac{2}{s+t^3}, \quad e^{\chi(t)} = 2 t^3,
    \label{eq:qhdc_schedule}
\end{equation}
where $s$ regularizes the kinetic prefactor at $t=0$. In the numerical simulations, we set $s=1$ and refer to this choice as the QHD-C schedule. Unless otherwise stated, the total evolution time is $T=10$, with time step $\delta t=10^{-3}$, corresponding to $10^4$ Trotter steps.

As a classical baseline, we use a fast-Fourier-transform-based split-operator solver for the discretized Schr\"odinger equation on the same lattice grid. The solver alternates between position and momentum representations using FFTs, which allows the potential and kinetic evolution operators to be applied efficiently in their diagonal bases. For the one-dimensional benchmark problems, we validate this solver against a direct SciPy ODE integration of the same semi-discrete Schr\"odinger equation with the same relaxation schedule and find agreement at the level of $10^{-9}$ or better.

\begin{figure*}[htbp]
    \centering
    \includegraphics[width=0.85 \linewidth]{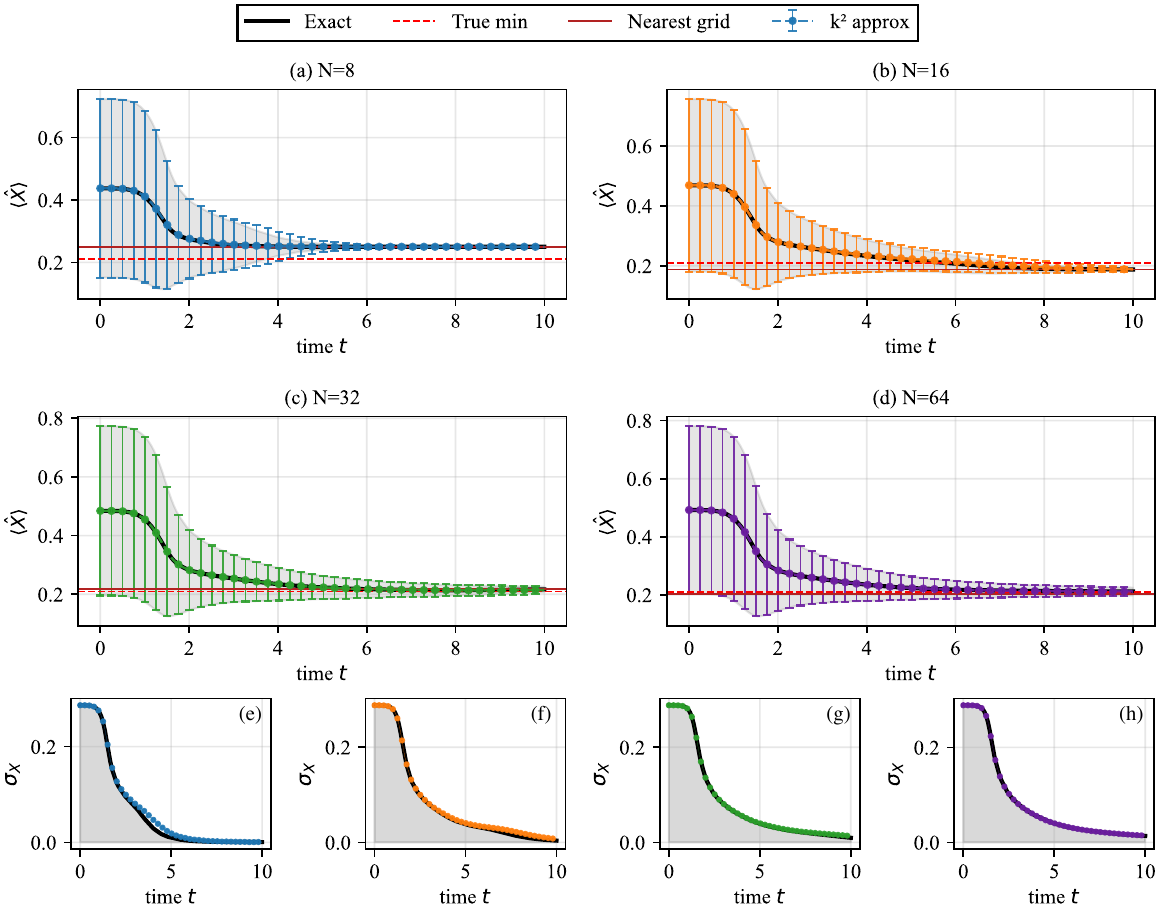}
    \caption{Comparison of the $k^2$-approximated kinetic operator with the exact dynamics as the number of grid points is varied. We use the two-mode cosine target function as the benchmark and solve the dynamics on (a) $N=8$, (b) $N=16$, (c) $N = 32$, and (d) $N = 64$ grid points. In these figures, the black curves show the exact classical dynamics of $\langle x\rangle$, the colored markers with error bars show the $k^2$-approximated circuit results, the red dashed lines indicate the true minimum, and the red solid lines indicate the nearest grid point to the minimum. (e) to (h) show the corresponding spatial fluctuations $\sigma_x$ for the same grid sizes. As the grid resolution increases, the deviation introduced by the $k^2$ approximation is reduced. We use periodic boundary conditions.}
    \label{fig:k2_vs_grid_points}
\end{figure*}

Fig.~\ref{fig:verify_circ} validates the one-hot and binary QHD circuit implementations on the four target objectives $f(x)$. For each encoding, the objective is mapped to the QHD potential and the initial state is chosen to be the uniform superposition over grid points,
\begin{equation}
    \ket{\psi_0} = \frac{1}{\sqrt{N}} \sum_{j=0}^{N-1} \ket{j}.
\end{equation}
After constructing the quantum circuit, we simulate it with a statevector simulator and extract the final state, denoted by $\ket{\psi(T)}$. We then compute the mean position $\mean{x}$ and the spatial fluctuation
\begin{equation}
    \sigma_x = \sqrt{\mean{x^2} - \mean{x}^2},
\end{equation}
where $\mean{\cdots}=\bra{\psi}\cdots\ket{\psi}$. Fig.~\ref{fig:verify_circ}a to~\ref{fig:verify_circ}d compare the classical split-operator solver with the one-hot and binary circuit simulations. The classical value of $\mean{x}$ is shown by the black curves, and the region $\mean{x}\pm\sigma_x$ is shown by the black shaded bands. The corresponding binary and one-hot circuit results are shown as blue and orange markers, respectively, with error bars indicating $\pm\sigma_x$ from the final simulated circuit states. In all four target cases, the one-hot and binary circuits agree well with the classical solver, and the QHD-C schedule drives the dynamics toward the optimal regions of the target objectives.

\subsection{Approximation in subroutines of kinetic evolution} \label{sec:numerics:approx}

The exact binary circuit implementation discussed in Sec.~\ref{sec:numerics:circuit_construction} gives a gate-level realization of the discretized Schr\"odinger equation used in the QHD dynamics. In that construction, the potential part is diagonal in the binary position basis, while the kinetic part is diagonal only after transforming each coordinate register to the Fourier basis. As shown in Fig.~\ref{fig:qhd_circs}, each binary kinetic block therefore consists of a QFT, a Fourier-basis diagonal phase operator, and an inverse QFT. The gate complexity reported for binary encoding in Table~\ref{tab:encoding_gate_scaling} has contributions from both the QFT layers and the diagonal phase synthesis. Since the exact implementation has already been verified against the reference split-operator dynamics, we now examine the low-momentum approximation to the kinetic eigenvalues and the AQFT truncation to the kinetic evolution.

\paragraph{Low-momentum kinetic-spectrum approximation.}
This approximation modifies only the Fourier-basis diagonal phase operator. The dominant cost within this exact kinetic block comes from synthesizing the diagonal phases associated with the finite-difference kinetic eigenvalues. These eigenvalues are trigonometric functions of the discrete momentum index, as shown in Sec.~\ref{sec:binary}, so exact phase synthesis generally produces many multi-qubit $Z$ rotations in the binary representation. To reduce this overhead, we replace the exact finite-difference spectrum by the low-momentum quadratic approximation proportional to $k^2$, as discussed in Sec.~\ref{sec:binary:comparison}. This keeps the same QFT-phase-inverse-QFT circuit structure, but converts the Fourier-basis kinetic phase into a quadratic function of the binary momentum bits, which can be synthesized using only one- and two-qubit $Z$ rotations.

This approximation is expected to be accurate when the evolving wavefunction has most of its support on low-momentum modes, where the finite-difference dispersion is well approximated by a quadratic dispersion. During QHD evolution, however, the time-dependent potential may induce higher-momentum components, which can lead to deviations from the exact lattice spectrum. To understand the size of this deviation, we test the dynamics on the benchmark one-dimensional target optimization problems and compare the results with the classical solver solutions.

We first isolate the effect of the low-momentum approximation by keeping the QFT exact and changing only the Fourier-basis kinetic phase, and hence separate the error contribution from the AQFT.
In Fig.~\ref{fig:verify_k2}, we compare the resulting $k^2$-based binary circuit dynamics with the classical split-operator solution for the same periodic one-dimensional benchmarks. Agreement in these observables indicates that the wavefunction remains sufficiently concentrated in the low-momentum sector for the quadratic kinetic approximation to reproduce the exact QHD dynamics, whereas visible deviations quantify the error introduced by replacing the exact finite-difference spectrum.

We also notice that the accuracy of the low-momentum approximation also depends on the grid resolution. As shown in Fig.~\ref{fig:k2_vs_grid_points}, the deviation from the exact dynamics decreases as the number of grid points is increased from $N=8$ to $N=64$. This behavior is expected because a finer grid gives a smaller lattice spacing and resolves the low-momentum part of the finite-difference spectrum more accurately. In this regime, the exact lattice dispersion approaches the continuum quadratic form, so replacing the finite-difference eigenvalues by the $k^2$ approximation introduces a smaller phase error during the kinetic evolution. Consequently, both the mean position $\mean{x}$ and the fluctuation $\sigma_x$ from the approximate circuit track the exact split-operator dynamics more closely at larger $N$.

\begin{figure}[htbp]
    \centering
    \subfloat[Fidelity to the classical solution]{%
        \includegraphics[width=0.8\linewidth]{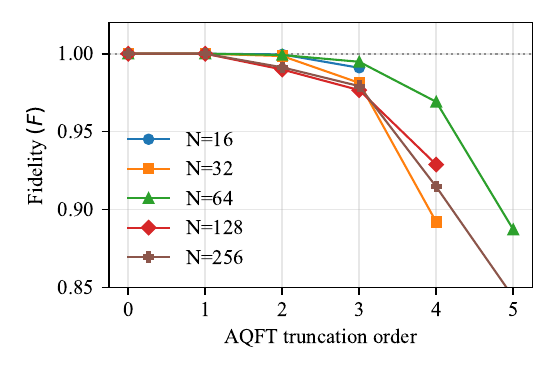}}%

    \subfloat[Success probability within $x_0 \pm 0.05$]{%
        \includegraphics[width=0.8\linewidth]{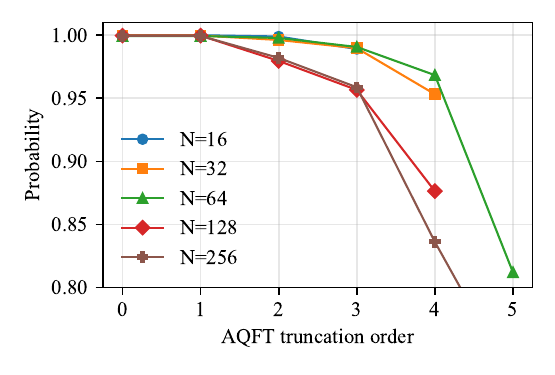}}%
    \caption{Effect of approximate-QFT truncation on the two-mode cosine one-dimensional periodic benchmark. Panel (a) shows the final-state fidelity relative to the reference evolution obtained from the classical Schr\"odinger-equation solver. Panel (b) shows the corresponding success probability, defined as the probability of measuring a point within $\pm 0.05$ of the optimal solution $x_0$, similar to the criterion used in Ref.~\cite{leng2023quantum}. For all spatial resolutions except $N=256$, we use $T=10$ with $10^5$ Trotter steps; for $N=256$, we use $2\times 10^5$ Trotter steps. The truncation order $d_{\mathrm{tr}}$ is defined such that controlled rotations with angles smaller than $2^{-(\log_2 N-d_{\mathrm{tr}})}$ are omitted.}
    \label{fig:aqft}
\end{figure}

\paragraph{AQFT.} We next consider the effect of the AQFT subroutine in the QHD solver. In the AQFT implementation, controlled rotations whose angles fall below a chosen cutoff are omitted, reducing the number of two-qubit controlled-phase gates in the QFT and inverse-QFT layers. This approximation is independent of the low-momentum approximation. One can use an exact QFT with the approximate kinetic spectrum, an AQFT with the exact finite-difference spectrum, or combine both approximations. Each choice changes the implemented unitary and can affect both observables, such as $\mean{x}$ and the solution quality of the QHD algorithm. As more controlled rotations are removed from the QFT, the final-state fidelity relative to the exact classical solver generally decreases. However, the QHD solution quality does not necessarily degrade at the same rate as the state fidelity, because optimization success depends mainly on whether probability remains concentrated near the minimizer.

To examine the effect of combining AQFT with the low-momentum approximation, in Fig.~\ref{fig:aqft}, we use the one-dimensional two-mode cosine target function as a benchmark. 
We evaluate the QHD dynamics with the AQFT-$k^2$ approximation by simulating the generated quantum circuits and compare the final-state fidelity with the classical Schr\"odinger-equation solver in Fig.~\ref{fig:aqft}a. We consider resolutions from $N=16$ to $N=256$ and observe that the state fidelity decreases as more QFT rotations are truncated for all resolutions. In Fig.~\ref{fig:aqft}b, we also compute the success probability, defined as the probability of the final state lying within $x_0\pm0.05$, consistent with the criterion used in Ref.~\cite{leng2023quantum}. For moderate truncation orders, the success probability decreases only mildly, from approximately $1$ to about $0.95$. The degradation becomes more abrupt for $N=256$, a trend that is also observed for the one-dimensional cosine target function (see Appendix).

These results indicate that AQFT truncation can change the full final state while still preserving the probability near the optimum over a range of truncation levels. In practical binary-encoded QHD applications, however, the leading kinetic complexity often comes from realizing the Fourier-basis eigenvalue phase rather than from the QFT layers alone (see Table~\ref{tab:encoding_gate_scaling}, and also the discussion in the following sections). Moreover, for practical optimization problems with a large number of variables, iterative refinement can keep the per-variable resolution moderate~\cite{wu2026ALQHD}. In such settings, the gate savings obtained by truncating controlled rotations in each AQFT subroutine may be limited because each variable is represented at relatively low resolution.

\section{Gate count complexity and fault-tolerant resource analysis} \label{sec:resource}

In Sec.~\ref{sec:binary:comparison}, we present asymptotic gate count scaling for the kinetic and potential circuit components of the QHD algorithm. While the kinetic term complexity is determined solely by the choice of encoding and the kinetic subroutine (QFT, AQFT, or low-moment approximation), the potential term realization depends on the analytic structure of the target function. The number of Pauli $Z$ terms in the diagonal decomposition, and hence the resulting gate count, varies with the separability and degree of the objective. Therefore, actual gate counts can differ substantially across optimization problem instances, even at the same grid resolution and encoding.

In this section, we quantify the gate-count scaling and FT resources required by the QHD circuit constructions. The goal is to connect the asymptotic resource estimates to concrete benchmark optimization problems and to compare the practical overheads of the one-hot and binary encodings across both the exact and approximated kinetic realizations.

\begin{figure}[htbp]
    \centering
    \includegraphics[width=0.95\linewidth]{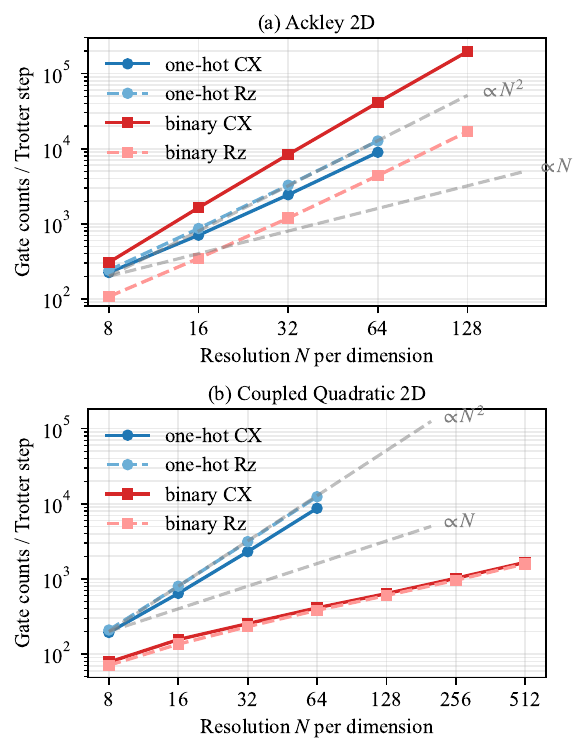}
    \caption{Total CX and $R_z$ gate counts for two benchmark problems: (a) Ackley 2D and (b) coupled quadratic 2D. The target functions are defined in App.~\ref{app:benchmark_targets}. The gray dashed lines indicate $N^2$ and $N$ scaling as visual guides and are not obtained from fits. Red curves correspond to the one-hot encoding, while blue curves correspond to the binary encoding with exact kinetic evolution. Because the generated one-hot circuits become large for resolutions $N>64$, we do not report those circuit sizes. For the same reason, we do not report the binary realization for the Ackley 2D target at resolutions $N>128$. Other parameters are $T=10$, $100$ total Trotter steps, and the QHD-C schedule with $s=1.0$.}
    \label{fig:rz_2d_scale}
\end{figure}

\subsection{Gate-count scaling for benchmark problems}

\begin{table*}[tbp]
\centering
\small
\setlength{\tabcolsep}{4.5pt}
\caption{Circuit statistics for one-hot and binary (exact) encoding across four 2D periodic benchmark targets at resolutions $N=32$ and $N=64$, using 100 Trotter steps with the QHD-C schedule. CNOT/step and $R_z$/step are gate counts per Trotter step. Total is the total gate count for the full 100-step circuit including CNOT, $R_z$, and all Clifford gates. G/D (gates per depth layer) measures circuit parallelism, where higher values indicate more operations that can be executed simultaneously. One-hot uses $2N$ qubits for a 2D problem while binary uses $2\log_2 N$ qubits ($64$/$128$ vs.\ $10$/$12$ at $N=32$/$64$, respectively). \textcolor{blue}{Blue} values indicate the encoding with the smaller gate count or larger G/D.}
\label{tab:depth_parallelism}
\begin{ruledtabular}
\begin{tabular}{l l l r r r r r}
  Resolution & Target & Encoding & CNOT/step & $R_z$/step & Total & Depth & G/D \\
  \hline
  \multirow{8}{*}{$N=32$}
  & \multirow{2}{*}{Ackley}
    & one-hot & \textcolor{blue}{$2{,}432$} & $3{,}264$ & $3{,}872{,}000$ & \textcolor{blue}{$143{,}400$} & \textcolor{blue}{$27.0$} \\
  & & binary  & $8{,}358$ & $1{,}189$ & $1{,}925{,}052$ & $1{,}780{,}504$ & $1.1$ \\[2pt]
  & \multirow{2}{*}{Alpine1}
    & one-hot & $384$ & $254$ & $958{,}200$ & \textcolor{blue}{$18{,}300$} & \textcolor{blue}{$52.4$} \\
  & & binary  & $360$ & $228$ & $260{,}352$ & $85{,}500$ & $3.1$ \\[2pt]
  & \multirow{2}{*}{Camel3}
    & one-hot & $2{,}306$ & $3{,}137$ & $3{,}745{,}100$ & $140{,}300$ & \textcolor{blue}{$26.7$} \\
  & & binary  & $332$ & $247$ & $274{,}652$ & $96{,}000$ & $2.9$ \\[2pt]
  & \multirow{2}{*}{Cpl.~Quad.}
    & one-hot & $2{,}306$ & $3{,}137$ & $3{,}745{,}100$ & $140{,}300$ & \textcolor{blue}{$26.7$} \\
  & & binary  & $254$ & $231$ & $252{,}452$ & $68{,}691$ & $3.7$ \\
  \hline
  \multirow{8}{*}{$N=64$}
  & \multirow{2}{*}{Ackley}
    & one-hot & \textcolor{blue}{$8{,}960$} & $12{,}672$ & $13{,}683{,}200$ & \textcolor{blue}{$271{,}400$} & \textcolor{blue}{$50.4$} \\
  & & binary  & $41{,}242$ & $4{,}387$ & $8{,}095{,}054$ & $7{,}855{,}638$ & $1.0$ \\[2pt]
  & \multirow{2}{*}{Alpine1}
    & one-hot & \textcolor{blue}{$768$} & $510$ & $1{,}918{,}200$ & \textcolor{blue}{$18{,}300$} & \textcolor{blue}{$104.8$} \\
  & & binary  & $796$ & $418$ & $478{,}354$ & $166{,}233$ & $2.9$ \\[2pt]
  & \multirow{2}{*}{Camel3}
    & one-hot & $8{,}706$ & $12{,}417$ & $13{,}428{,}300$ & $268{,}300$ & \textcolor{blue}{$50.1$} \\
  & & binary  & $640$ & $424$ & $468{,}154$ & $178{,}833$ & $2.6$ \\[2pt]
  & \multirow{2}{*}{Cpl.~Quad.}
    & one-hot & $8{,}706$ & $12{,}417$ & $13{,}428{,}300$ & $268{,}300$ & \textcolor{blue}{$50.1$} \\
  & & binary  & $412$ & $382$ & $407{,}554$ & $110{,}015$ & $3.7$ \\
\end{tabular}
\end{ruledtabular}
\end{table*}

We consider both one-dimensional target functions that are considered in Sec.~\ref{sec:circuits} and four additional two-dimensional target functions (see Appendix~\ref{app:benchmark_targets}) as benchmarking problems, so that the resource estimates capture the dependence on grid resolution as well as on the number of optimization variables. For each benchmark, we generate the corresponding one-hot and binary-encoded QHD circuits using the potential and kinetic implementations described and verified in Sec.~\ref{sec:circuits}. 

For the QHD evolution, we use the same QHD-C schedule defined in Eq.~\eqref{eq:qhdc_schedule} and tested in the previous section, and keep the total evolution time fixed at $T=10$. Since the purpose of this section is to analyze gate counts and resource scaling rather than to further validate the dynamical accuracy, we use $100$ total Trotter steps for the benchmark circuits. This choice is sufficient for resource estimation because the circuit structure of the QHD algorithm is repeated from one Trotter step to the next, with only the time-dependent rotation angles changing along the schedule. After constructing and verifying these circuits, we transpile them into a Clifford+$R_z$ gate set using qiskit built-in transpiler with optimization level \texttt{0}. We further set an angle threshold $\delta\theta$ for the rotation gates. If a rotation angle is within this threshold of an $S$, $S^\dagger$, Pauli-$Z$, or identity gates, we replace the rotation by the corresponding Clifford gate and remove it from the $R_z$ gate count.

In Fig.~\ref{fig:rz_2d_scale}, we plot the CNOT and Rz gate counts for QHD algorithms for solving two 2D benchmarking example problems. In these plots, we transpiled the circuits as mentioned above. More benchmarking problems with different resolutions can be found in Appendix~\ref{app:gate_scaling}. We noticed that in the Ackley example, both binary and one-hot realization gives a $R_z$ and CX scalings close to $N^2$, while in the coupled quadratic 2D example, the binary realization shows a better scaling. Furthermore, in the Ackley example, the binary realization uses more CNOT than one-hot, while saving the number of $R_z$ gates. But in the coupled quadratic 2D example, binary encoding has fewer CNOT gates. 

The distinct scaling behaviors arise from the structure of the potential evolution. In the binary encoding, the potential is implemented via a Walsh-Hadamard transform (WHT) of the objective values at each computational basis state, which decomposes the diagonal operator into a sum of multi-qubit Pauli-$Z$ rotations. Therefore, the complexity of the required entangling gates and decomposed rotation gates depends on the analytic structure of the target function.

For the Coupled Quadratic example, the target function is defined as
\begin{align}
    f(x,y) &=(x-0.25)^2+1.4(y-0.65)^2 \nonumber \\
    &\quad +0.2(x-0.25)(y-0.65).
\end{align}
This target function has a polynomial structure of degree at most two, which limits the Pauli-$Z$ rotations in the binary decomposition to weight at most two. This structure leads to the substantially smaller and more slowly growing CNOT count in Fig.~\ref{fig:rz_2d_scale}b. In contrast, the Ackley target function is
\begin{align}
    f(x,y) &=-20\exp\left(-0.2\sqrt{\frac{X^2+Y^2}{2}}\right) \nonumber \\
    &-\exp\left(\frac{\cos(2\pi X)+\cos(2\pi Y)}{2}\right)
    +20+\mathrm{e},
\end{align}
where the term such as $\exp(-0.2\sqrt{0.5(X^2+Y^2)})$ couples both variables through a single transcendental function. This coupling generally produces nontrivial support across high-weight Pauli-$Z$ rotations on the full set of $2n$ binary qubits, where $n=\log_2 N$ is the number of qubits per variable. The resulting dense WHT decomposition causes the CX count to scale as $O(N^2\log N)$, consistent with the observed $\sim N^2$ growth in Fig.~\ref{fig:rz_2d_scale}a. In the one-hot encoding, however, this distinction is less pronounced because the potential cost is determined by the number of grid-point pairs appearing in cross-variable terms. The CNOT count therefore scales as $O(N^2)$ for coupled functions, largely independent of the analytic form of the function. Thus, one-hot encoding provides more uniform cost scaling across different target functions, but it does not exploit the structural sparsity that makes polynomial objectives inexpensive under binary encoding. 

\begin{figure}[htbp]
    \centering
    \includegraphics[width=0.8\linewidth]{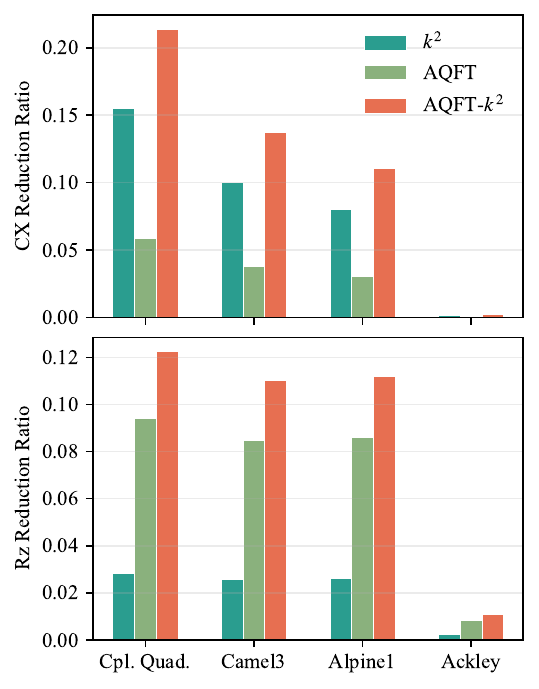}
     \caption{The CNOT and $R_z$ gate counts reduction relative to the exact dynamics for four 2D benchmark problems using different kinetic-evolution approximations with resolution $N=64$. The CNOT and $R_z$ gate count reduction ratios are shown in the upper and lower panels, respectively. We compare the low-momentum approximation with exact QFT (the $k^2$ curve), exact momentum evolution with AQFT (the AQFT curve), and AQFT combined with the low-momentum approximation (the AQFT-$k^2$ curve) against the exact dynamics. All other parameters are identical to those used in Fig.~\ref{fig:rz_2d_scale}.}
    \label{fig:all4_kinetic}
\end{figure}

In Table~\ref{tab:depth_parallelism}, we further compare circuit depth and parallelism for the two encodings across all four 2D benchmark targets at resolutions $N=32$ and $N=64$. Several consistent trends are visible across all four targets. First, binary encoding achieves fewer $R_z$ gates per Trotter step than one-hot in every case. In binary encoding, each multi-qubit Pauli-$Z$ term requires exactly one $R_z$ in its CNOT ladder, whereas one-hot applies one $R_z$ per grid-point pair for coupled terms, growing as $O(N^2)$. This $R_z$ advantage translates directly into a lower $T$-gate count in fault-tolerant compilation (Sec.~\ref{sec:ft_resources}). Second, the CNOT comparison reverses depending on the target functions. The binary encoding realization is far cheaper for polynomial objective functions (Coupled Quadratic, Camel3), where the WHT is sparse, but more expensive than one-hot for Ackley, where the $O(N^2)$ expansion of the potential evolution dominates.

The G/D column computes the average number of gates per depth of the quantum circuits, which reveals a structural parallelism contrast that holds uniformly across all targets. One-hot achieves G/D values of $27$ to $105$, versus $1.0$ to $3.7$ for binary, which is a $15$ to $50\times$ better. This arises because the potential Pauli-$Z$ operators in one-hot act on disjoint qubit pairs and can be executed in parallel, while binary's multi-qubit Pauli chains share register qubits and require sequential CNOT ladders. Despite this advantage in parallelism ratio, one-hot does not always achieve a shallower circuit in absolute terms. For polynomial targets (Coupled Quadratic, Camel3), the total binary gate count is so small that binary achieves both fewer gates and shallower circuit depth. For Ackley at $N=64$, the binary circuit depth exceeds $7.8\times10^6$ layers, while one-hot remains $\sim 29\times$ shallower at $\sim 2.71 \times 10^5$ layers, at the cost of using $10$ times more qubits.

\begin{figure}[tbp]
    \centering
    \includegraphics[width=0.8\linewidth]{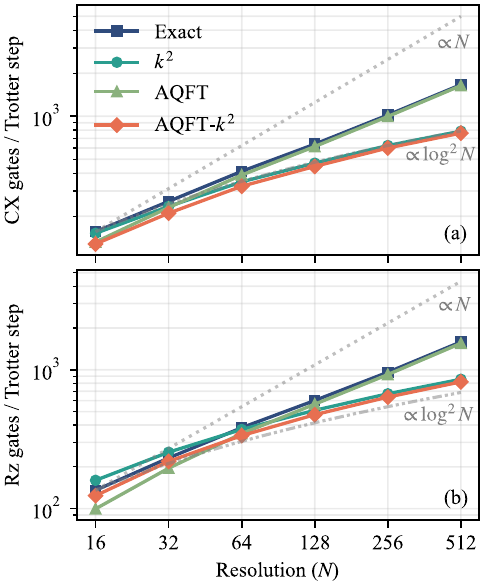}
    \caption{Gate-count scaling for different kinetic-evolution approximations for Coupled Quadratic target function. The CNOT and $R_z$ gate counts per Trotter step are shown in (a) and (b), respectively. The other settings are identical to Fig.~\ref{fig:all4_kinetic}.}
    \label{fig:cq_kinetic}
\end{figure}

The low-momentum approximation and the AQFT subroutine can reduce the circuit cost of realizing the kinetic terms in QHD dynamics. In Fig.~\ref{fig:all4_kinetic}, we show results for all four 2D benchmark problems using the low-momentum approximation with exact QFT (the $k^2$ curve), exact momentum evolution with AQFT (the AQFT curve), and AQFT combined with the low-momentum approximation (the AQFT-$k^2$ curve), each compared against the exact dynamics. For the Coupled Quadratic, Camel3, and Alpine1 cases, both approximations reduce the $R_z$ gate counts $10\%$ to $12\%$, and reduce the CNOT gate counts by $10\%$ to $20\%$. For the Ackley case, however, the reduction is much less pronounced. As discussed above, the Ackley circuit is dominated by the more expensive potential evolution realization, so savings from the kinetic realization provide only marginal benefits. In this regime, both the CNOT and $R_z$ gate counts are dominated by the potential evolution.

To further understand the scaling behavior and crossover regimes of approximate kinetic-evolution methods, we plot the gate-count scaling with resolution $N$ for the Coupled Quadratic 2D benchmark in Fig.~\ref{fig:cq_kinetic}. The other three benchmarks can be found in Appendix~\ref{app:gate_scaling}. Because we fix the AQFT truncation order to $3$, AQFT alone provides only limited reductions in the CNOT and $R_z$ counts over the tested resolution range. By contrast, the low-momentum approximation (the $k^2$ curves) becomes important for reducing the $R_z$ gate count to $\sim \log^2(N)$, consistent with the scaling analysis in Sec.~\ref{sec:binary:comparison}. Therefore, we conclude that in practical QHD implementations, the low-momentum approximation is especially useful when the number of optimization variables is large but the required grid resolution for each variable is moderate.

\subsection{Fault-tolerant resource model and analysis} \label{sec:ft_resources}

The interpretation of the gate counts depends on the computational regime. For near-term, especially NISQ devices, the resource bottleneck is usually set by the physical native entangling gates of the hardware. Single-qubit rotations, including virtual or frame-updated $R_z$ gates, can often be implemented with low overhead, whereas entangling gates typically require calibrated two-qubit interactions and contribute more strongly to circuit duration and physical error rates~\cite{McKay2017}. The two-qubit Clifford gate counts within the Clifford+$R_z$ circuits above are therefore useful for identifying the circuit-level entangling overhead before imposing a fault-tolerant cost model.

In the FT setting, the cost hierarchy changes as discussed in Sec.~\ref{sec:one-hot:res-model}. We use a surface-code-based model in which logical Clifford operations, Pauli measurements, and lattice-surgery-type entangling operations are treated as lower-cost operations than non-Clifford gates, although they still consume code cycles and logical-qubit patches~\cite{Horsman2012,Brown2017,Litinski2019}. Universal quantum computation requires operations beyond the Clifford group, since Clifford stabilizer circuits alone are efficiently classically simulable~\cite{gottesman1998, Aaronson2004}, and the Eastin-Knill theorem rules out a universal set of transversal logical gates for a QEC code~\cite{Eastin2009}. In surface-code architectures, the resulting non-Clifford resources, conventionally counted as $T$ gates or magic states, therefore become major cost drivers in fault-tolerant resource estimates~\cite{litinski2019magic,gidney2024cultivation}.

\paragraph{Clifford+$T$ transpilation. }

To give a better resource analysis on the QHD algorithm applied to the benchmarking problems, we use \texttt{NWQEC} to transpile the Clifford+$R_z$ circuit descriptions into Clifford+$T$ circuits~\cite{nwqec_toolkit,wang2024optimizing, wang2025tableau}. Before synthesis, each input circuit undergoes a lightweight optimization pass that removes locally canceling gate sequences. \texttt{NWQEC} then applies its standard preprocessing and $R_z$-cleanup passes, removing identity rotations and replacing $R_z$ rotations that are equivalent to $Z$, $S$, $S^\dagger$, $T$, and $T^\dagger$ by the corresponding gates. Each remaining arbitrary $R_z$ rotation is approximated by an ancilla-free Clifford+$T$ sequence using the GridSynth implementation of the Ross-Selinger synthesis algorithm~\cite{ross2014optimal}.

We target a total simulation error below $2\times 10^{-2}$ for the full QHD circuits~\footnote{This threshold is motivated by our numerical observations: when the average fidelity relative to the classical solver is above $0.98$, the success probability, defined as the probability of measuring a point within $0.1$ of the optimal solution $x_0$, remains nearly unchanged, with differences below $10^{-4}$.}. For the numerical verification, we use $10^4$ Trotter steps for all tested cases and resolutions. This choice is based on verification runs up to resolution $N=128$. Guided by state-dependent error analyses for time-dependent Hamiltonian simulation~\cite{an2021timeDependentUnbounded}, where the required step count can have weak or even $O(1)$ dependence on the grid resolution, we use the same Trotter-step count for larger resolutions in the resource analysis.

We allocate separate error budgets to $R_z$-synthesis errors and stochastic FT errors, including QEC failures and $T$-state preparation errors. For FT transpilation, we generate circuits using the same QHD-C schedule with $100$ Trotter steps, which keeps the transpilation problem tractable while preserving the repeated per-step circuit structure. We therefore set the total $R_z$-synthesis error budget to $\epsilon_{\mathrm{syn}}=10^{-4}$ and distribute this budget uniformly over the remaining $N_{R_z}$ rotations. The synthesis tolerance assigned to each rotation is $\epsilon_{R_z} = \epsilon_{\mathrm{syn}} / N_{R_z}$. This budget accounts only for the approximation error introduced by rotation synthesis and is separate from the logical failure probability associated with quantum error correction. The transpilation produces circuits containing only Clifford and $T$ gates.

\paragraph{Fault-tolerant resource estimation.}
The resource analysis uses \texttt{NWQRE}, a lightweight resource-estimation tool developed by the authors~\cite{song2026nwqre}. We estimate the FT resources using the surface-code procedure introduced in Ref.~\cite{Litinski2019}, assuming a physical error rate of $p_{\mathrm{phys}} = 10^{-4}$. The Clifford+$T$ circuit is first rewritten in a Pauli-based-computation representation. Clifford operations are commuted through the circuit, combined or canceled where possible, and ultimately absorbed into the final Pauli-product measurements~\cite{wang2025tableau}. The remaining computational sequence consists of non-Clifford $\pi/8$ Pauli-product rotations and final Pauli measurements.

Each non-Clifford rotation is implemented by consuming one distilled $T$ magic state. We use the fast data-block layout, which can consume one magic state and execute one non-Clifford rotation per surface-code time step. One such time step corresponds approximately to $d$ surface-code cycles for code distance $d$. Sufficient distillation blocks are provisioned so that magic-state production can sustain this consumption rate and does not become the computational bottleneck.

We evenly split the error budget for logical failure between QEC failure and $T$-state preparation. The magic-state distillation protocol and, when required, the number of distillation levels are selected so that the aggregate error contribution from all consumed magic states is below $\epsilon_T = 5\times10^{-5}$.
Thus, for a circuit containing $N_T$ non-Clifford rotations and distilled magic states with output error probability $p_T^{\mathrm{out}}$, the selected distillation configuration satisfies
\begin{equation}
N_T \cdot  p_T^{\mathrm{out}} < \epsilon_T.
\end{equation}

Independently, the surface-code distance is selected so that the aggregate probability of a logical failure during the computation is below $\epsilon_{\mathrm{L}} = 5\times10^{-5}$. Following the logical-error model used in Ref.~\cite{Litinski2019}, the logical error probability per logical qubit per code cycle is approximated as
\begin{equation}
p_{\mathrm{L}}(p_{\mathrm{phys}},d)
\approx
0.1\left(100p_{\mathrm{phys}}\right)^{(d+1)/2}.
\end{equation}
The smallest code distance satisfying
\begin{equation}
\mathcal{V}_{\mathrm{L}} \cdot
p_{\mathrm{L}}(p_{\mathrm{phys}},d)
< \epsilon_{\mathrm{L}}
\end{equation}
is used, where $\mathcal{V}_{\mathrm{L}}$ is the total logical space-time volume measured in logical-qubit code cycles, including the data block, routing and ancillary regions, magic-state factories, and magic-state storage. The two stochastic fault-tolerance budgets satisfy
$\epsilon_T+\epsilon_{\mathrm{L}}=10^{-4}$, separately from the approximation error introduced during $R_z$ synthesis. The selected code distance, data-block size, distillation configuration, and number of surface-code time steps are then used to determine the physical-qubit footprint and execution time of each benchmark circuit.

\begin{figure}[tbp]
  \centering
  \includegraphics[width=0.9\linewidth]{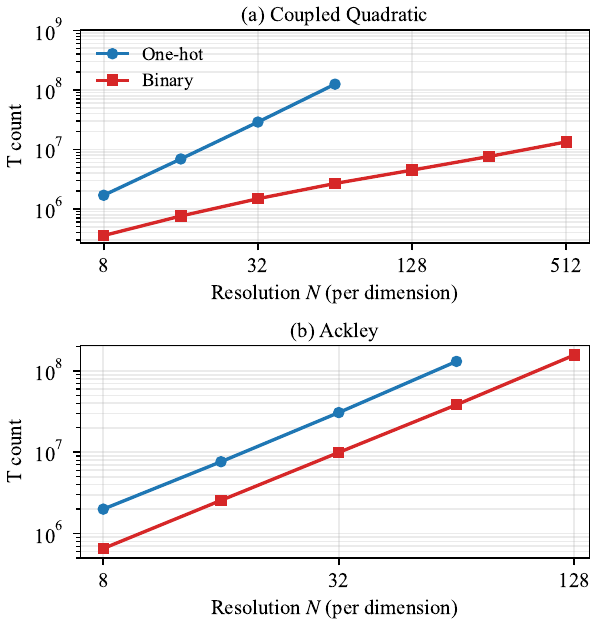}
  \caption{Total $T$-gate counts for two benchmark problems: (a) coupled quadratic 2D and (b) Ackley 2D. Red curves correspond to one-hot encoding, while blue curves correspond to binary encoding with exact kinetic evolution. The fault-tolerant decomposition and transpilation settings are described in the main text.}
  \label{fig:t_onehot_binary}
\end{figure}

Fig.~\ref{fig:t_onehot_binary} shows the $T$-gate counts for the coupled-quadratic and Ackley benchmarks. Consistent with the $R_z$ gate-count analysis in Fig.~\ref{fig:all4_kinetic}, binary encoding generates fewer $R_z$ rotations in both cases and therefore requires fewer synthesized $T$ gates. For Ackley, most of the $R_z$ rotations arise from the potential term, whose scaling is similar for one-hot and binary encodings. Consequently, the corresponding $T$-gate counts have similar scaling with resolution. For the coupled-quadratic benchmark, by contrast, the binary potential decomposition is much sparser, leading to a different and more favorable $R_z$ and $T$-gate scaling with resolution.

\begin{figure}[htbp]
  \centering
  \includegraphics[width=0.9\linewidth]{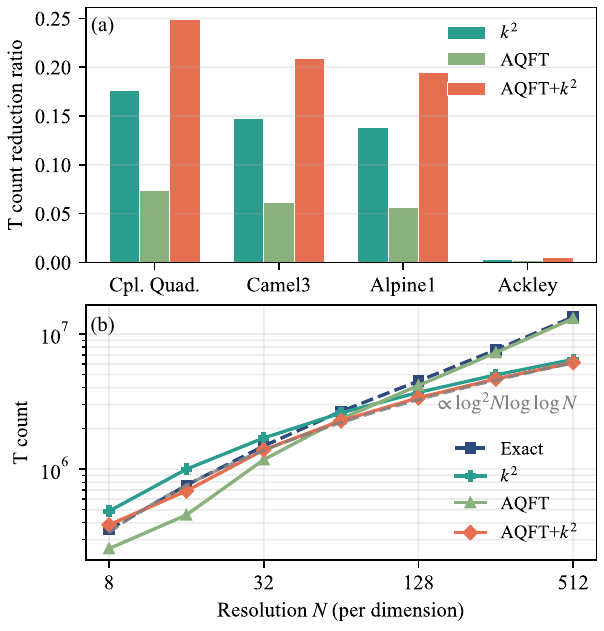}
  \caption{$T$-gate count reduction and scaling for fault-tolerant QHD implementations using approximate kinetic-evolution methods. Panel (a) shows the $T$-gate count reduction relative to the exact binary-encoding realization at resolution $N=512$. Panel (b) shows the total $T$ counts for different approximate kinetic-evolution methods applied to the coupled-quadratic benchmark problem. The gray dashed line indicates the scaling $\log^2(N)\log[\log(N)]$ as a visual guide.}
  \label{fig:t_gate}
\end{figure}

To further quantify the benefits of kinetic-term approximations, Fig.~\ref{fig:t_gate}a compares the $T$-gate count reductions obtained from the low-momentum approximation, the AQFT subroutine, and their combination, all relative to the exact binary-encoding dynamics. Similar to the $R_z$ gate-count trends, the low-momentum approximation gives the larger reduction in $T$ gates. For the coupled-quadratic, Three-Hump Camel (Camel3), and Alpine1 benchmarks, the total $T$-gate count is reduced by approximately $10\%$-$25\%$, and at resolution $N=512$ the reduction can reach $30\%$-$50\%$. In contrast, the Ackley benchmark shows only a marginal reduction, consistent with the $R_z$ analysis in Fig.~\ref{fig:all4_kinetic}. For the Ackley target, the dominant resource cost comes from the potential evolution. Fig.~\ref{fig:t_gate}b shows the $T$-gate counts for the coupled-quadratic benchmark, where the potential realization is not the dominant source of $R_z$ rotations. As the resolution increases, the kinetic approximation therefore produces a more significant reduction in total $T$ count. The numerically obtained $T$-gate counts also follow the predicted scaling $\sim \log^2(N)\log[\log(N)]$, shown by the gray dashed curve as a visual guide.

\begin{table*}[tbp]
\centering
\small
\setlength{\tabcolsep}{4.5pt}
\caption{Fault-tolerant resource comparison between one-hot and binary (exact) encoding for four 2D benchmark problems at resolutions $N=32$ and $N=64$. $T$ is the total number of $T$ gates after Clifford+$T$ transpilation (Sec.~\ref{sec:ft_resources}). $\Delta T$ is the percentage reduction in $T$-gate count relative to the one-hot row for the same target and resolution. $d$ is the surface-code distance required to meet the target logical error budget. $N_q$ is the number of physical qubits allocated to the data block, and $\Delta N_q$ is its percentage reduction relative to one-hot.}
\label{tab:ft_resource_reduction}
\begin{ruledtabular}
\begin{tabular}{l l l c c c c c}
& Target & Enc. & $T$ & $\Delta T$ & $d$ & $N_q$ & $\Delta N_q$ \\
\hline
\multirow{8}{*}{$N=32$}
& \multirow{2}{*}{Ackley}
  & one-hot & $3.08\times10^{7}$ & -- & $15$ & $6.84\times10^{4}$ & -- \\
& & binary & $9.95\times10^{6}$ & $-67.7\%$ & $13$ & $1.01\times10^{4}$ & $-85.2\%$ \\[2pt]
& \multirow{2}{*}{Alpine1}
  & one-hot & $2.13\times10^{6}$ & -- & $13$ & $5.14\times10^{4}$ & -- \\
& & binary & $1.48\times10^{6}$ & $-30.8\%$ & $13$ & $1.01\times10^{4}$ & $-80.3\%$ \\[2pt]
& \multirow{2}{*}{Camel3}
  & one-hot & $2.95\times10^{7}$ & -- & $15$ & $6.84\times10^{4}$ & -- \\
& & binary & $1.63\times10^{6}$ & $-94.5\%$ & $13$ & $1.01\times10^{4}$ & $-85.2\%$ \\[2pt]
& \multirow{2}{*}{Cpl.~Quad.}
  & one-hot & $2.88\times10^{7}$ & -- & $15$ & $6.84\times10^{4}$ & -- \\
& & binary & $1.48\times10^{6}$ & $-94.9\%$ & $13$ & $1.01\times10^{4}$ & $-85.2\%$ \\
\hline
\multirow{8}{*}{$N=64$}
& \multirow{2}{*}{Ackley}
  & one-hot & $1.31\times10^{8}$ & -- & $15$ & $1.30\times10^{5}$ & -- \\
& & binary & $3.84\times10^{7}$ & $-70.7\%$ & $13$ & $1.18\times10^{4}$ & $-90.9\%$ \\[2pt]
& \multirow{2}{*}{Alpine1}
  & one-hot & $4.44\times10^{6}$ & -- & $13$ & $9.77\times10^{4}$ & -- \\
& & binary & $3.02\times10^{6}$ & $-31.9\%$ & $13$ & $1.18\times10^{4}$ & $-87.9\%$ \\[2pt]
& \multirow{2}{*}{Camel3}
  & one-hot & $1.28\times10^{8}$ & -- & $15$ & $1.30\times10^{5}$ & -- \\
& & binary & $3.06\times10^{6}$ & $-97.6\%$ & $13$ & $1.18\times10^{4}$ & $-90.9\%$ \\[2pt]
& \multirow{2}{*}{Cpl.~Quad.}
  & one-hot & $1.25\times10^{8}$ & -- & $15$ & $1.30\times10^{5}$ & -- \\
& & binary & $2.68\times10^{6}$ & $-97.9\%$ & $13$ & $1.18\times10^{4}$ & $-90.9\%$ \\
\end{tabular}
\end{ruledtabular}
\end{table*}

Table~\ref{tab:ft_resource_reduction} summarizes the fault-tolerant resource footprint of one-hot and binary (exact) encoding for all four 2D benchmark problems at $N=32$ and $N=64$. The qubit savings of binary encoding established in Sec.~\ref{sec:ft_resources} for the logical-qubit count, which scales as $O(\log N)$ rather than $O(N)$, are only part of the advantage realized at the fault-tolerant level. The table shows that binary encoding also reduces the total $T$-gate count relative to one-hot in every benchmark, by as little as $31\%$ for Alpine1 up to nearly $98\%$ for the coupled-quadratic and Three-Hump Camel targets, consistent with the trends already observed in Fig.~\ref{fig:t_onehot_binary}.

This reduction in $T$-gate count is significant because it directly lowers the burden placed on the surrounding quantum error correction. Each $T$ gate consumes a magic state and one code-cycle time step in the fast data-block layout (Sec.~\ref{sec:ft_resources}), so a circuit with fewer $T$ gates also requires fewer total code cycles and hence a smaller logical space-time volume $\mathcal{V}_{\mathrm{L}}$. Since the required code distance $d$ is set by the condition that $\mathcal{V}_{\mathrm{L}}\,p_{\mathrm{L}}(p_{\mathrm{phys}},d)$ remains below the logical-error budget $\epsilon_{\mathrm{L}}$, a smaller $\mathcal{V}_{\mathrm{L}}$ allows the same error budget to be met at a smaller $d$. This effect is also visible directly in Table~\ref{tab:ft_resource_reduction}. For Ackley, Three-Hump Camel, and the coupled-quadratic target, the one-hot circuits accumulate enough $T$ gates and code cycles to require $d=15$, whereas the binary circuits, with $68\%$ to $98\%$ fewer $T$ gates, only require $d=13$ at the same resolutions. Because the physical-qubit footprint of a logical qubit patch scales approximately as $d^2$, this reduction in code distance compounds with the reduction in logical-qubit count to produce the much larger reduction in physical data qubits, reaching $85\%$ to $91\%$ across the four benchmarks. For Alpine1, by contrast, the one-hot circuit already has few enough $T$ gates to qualify for $d=13$, so both encodings share the same code distance at this resolution. However, as we have shown in the Appendix~\ref{app:gate_scaling}, as we increase the resolution, the binary encoding implementations tend to have more $T$ gate savings. But in the example resolution shown in Table~\ref{tab:ft_resource_reduction}, the physical-qubit savings come solely from the reduced logical-qubit count of binary encoding, which shows that the $T$-count-driven reduction in code distance and the encoding-driven reduction in logical-qubit count are two distinct and complementary mechanisms by which binary encoding lowers the FT resource cost of QHD circuits.

\section{Discussion}
\label{sec:discussion}

In the analytical gate scaling analysis, we count the cost of a single Trotter step. If the number of steps is also included using the standard worst-case analysis with operator-norm error bound, then a time-independent one-dimensional discretized Schr\"odinger Hamiltonian $H=A+B$ requires a step count that grows linearly with the number of spatial grid points. In particular, the second-order Trotter estimate scales as $(\|[A,[A,B]]\|+\|[B,[B,A]]\|)^{1/2}$, and for real-space Schr\"odinger discretizations with a smooth bounded potential these nested commutators scale as $\|[A,[A,B]]\|=\mathcal{O}(N^2)$ and $\|[B,[B,A]]\|=\mathcal{O}(N)$~\cite{an2021timeDependentUnbounded,childs2021theory}. Thus, at fixed final time and target error, the conservative operator-norm estimate gives $r=\mathcal{O}(N)$. This factor does not change the per-step gate counts, but it multiplies the total number of circuit rotations. If arbitrary $R_z$ rotations are synthesized under a fixed total error budget, the larger rotation count can also require higher per-rotation precision, adding a logarithmic factor in $N$ to the $T$-gate synthesis cost. It should be interpreted as a conservative worst-case correction. For sufficiently smooth or low-momentum states, vector-norm errors can show weaker dependence on $N$.

Even with binary encoding, a fully digital QHD implementation requires one position register for each optimization variable. For $d$ variables and $N$ grid points per variable, the data-register cost therefore scales as $d\log_2 N$ qubits. This is exponentially smaller than storing the full classical grid, whose size scales as $N^d$, but the qubit count still grows linearly with the optimization dimension. As a result, high-dimensional problems can remain challenging for near-term early-FT implementations. Iterative refinement can reduce this cost by keeping the per-variable grid resolution moderate at each refinement stage~\cite{wu2026ALQHD}, but it does not remove the linear dependence on the number of variables.

The circuit constructions studied in this work can also be viewed as first-quantized simulations of an engineered time-dependent Schr\"odinger Hamiltonian on the optimization landscape. This perspective makes the connection to existing first-quantized quantum-simulation methods explicit. The potential term is diagonal in the position basis, while the kinetic term is implemented through either local hopping terms or a transform-basis phase operator. The main distinction is that the configuration-space dimension is set by the number of optimization variables rather than by the physical phase space dimension. Consequently, techniques developed for first-quantized simulation may be useful for improving QHD implementations, but their effectiveness must be assessed in the higher-dimensional optimization setting.

The numerical results also depend on the choice of the QHD schedule. In this work, the schedule is treated as an input to the circuit construction rather than as an optimized control function. Different schedules can change the amount of diabatic excitation, the momentum content generated during the evolution, and the number of Trotter steps required to reach a given solution quality. Developing principled schedule-design or schedule-optimization strategies is therefore an important direction for future work.

Finally, the resource estimates reported here do not fully optimize the FT implementation of the Clifford+$R_z$ circuits over ancillary qubits. Because non-Clifford resources are often more expensive than additional logical qubit patches in surface-code architectures, ancilla-assisted rotation synthesis, phase-gradient methods, or other space-time tradeoffs could reduce the effective $T$ count or $T$ depth. Incorporating such optimizations would refine the quantitative resource estimates, although it would not change the encoding-level scaling comparisons emphasized in this work.

\section{Conclusion}
\label{sec:conclusion}

We have developed an encoding-aware circuit and resource analysis for digital implementations of the QHD algorithm. Starting from the same discretized QHD dynamics, we derived and compared gate-count scalings for coordinate-wise one-hot encoding and binary amplitude encoding, constructed explicit Clifford+$R_z$ circuits for both approaches, validated the resulting dynamics against classical Schr\"odinger-equation solvers, and evaluated benchmark gate counts and FT Clifford+$T$ resource estimates. This provides a unified workflow for connecting the structure of a continuous optimization objective to the logical resources required by a digital QHD implementation.

With our study, we find that the benefits of adopting binary encoding are not only the reduction of qubit counts. Binary encoding reduces the data register from $O(dN)$ to $O(d\log N)$ qubits and can achieve comparable or better non-Clifford rotation scaling than one-hot encoding. In particular, when the binary kinetic evolution is combined with a low-momentum approximation and AQFT subroutines, the kinetic cost can be further reduced from the generic exact-QFT scaling to polylogarithmic scaling in the grid resolution. The numerical tests show that the low-momentum approximation can reproduce the relevant QHD observables well when the dynamics remains concentrated in the low-momentum sector, and that moderate AQFT truncation may reduce circuit cost while preserving high success probability near the optimizer even when the final-state fidelity is reduced.

At the same time, the benchmark results show that the potential term is often the dominant bottleneck. For polynomial or otherwise structured objectives, such as the coupled quadratic and three-hump camel examples, the binary Walsh-Hadamard transform gives a sparse Pauli-$Z$ decomposition and binary encoding gives substantially smaller gate counts and depths. For dense transcendental objectives such as Ackley, the binary potential decomposition produces high-weight Pauli-$Z$ rotations over many register qubits, so the savings from the kinetic term can be largely erased. In this regime, one-hot encoding uses many more qubits but can provide more uniform scaling, lower-weight potential rotations, better parallelism, and in some cases, circuits with smaller depth. These observations suggest that practical QHD compilation should account for objective structure. Binary encoding is attractive for sparse, low-degree, separable, or weakly coupled potentials, while one-hot or hybrid encodings may remain useful when dense potential synthesis or hardware connectivity dominates the cost.

The present analysis also has several limitations. The resource estimates are based on specific product-formula circuits, QHD schedules, synthesis choices, and benchmark objectives, and they do not fully optimize over ancillary qubits, phase-gradient methods, lattice-surgery layouts, or other space-time tradeoffs. The accuracy of the kinetic approximations depends on the momentum content generated by the schedule and objective landscape, and the total Trotter-step count may introduce additional resolution-dependent overhead beyond the per-step estimates emphasized here. Finally, although binary encoding reduces the dependence on grid resolution per variable, the number of registers still grows linearly with the optimization dimension, so high-dimensional applications will require additional algorithmic structure, decomposition, or iterative refinement.

These results indicate that fault-tolerant QHD should be designed through joint choices of encoding, potential compilation backend, kinetic approximation, schedule, and hardware layout. Future work should optimize QHD schedules together with circuit resources, develop structure-exploiting diagonal synthesis methods for broader objective classes, quantify Trotter and approximation errors in higher-dimensional settings, and explore hybrid or adaptive encodings that combine the qubit efficiency of binary representations with the locality and parallelism advantages of one-hot constructions.

\begin{acknowledgments}
The quantum resource analysis and the quantum algorithm development in this research is supported by the U.S. Department of Energy, Office of Science, National Quantum Information Science Research Centers, Codesign Center for Quantum Advantage (C2QA) under contract number DE-SC0012704, (Basic Energy Sciences, PNNL FWP 76274).
The algorithm software and resource analysis tool used in this research is developed and supported by Pacific Northwest National Laboratory’s Quantum Algorithms and Architecture for Domain Science (QuAADS) Laboratory Directed Research and Development (LDRD) Initiative. 
M.L., M.Z, and Y.C are supported under the QuAADS LDRD initiative. 
This research used resources of the National Energy Research Scientific Computing Center (NERSC), a DOE Office of Science User Facility supported by the Office of Science of the U.S. Department of Energy under Contract No. DE-AC02-05CH11231 using NERSC award  NERSC DDR-ERCAP0038957.
The Pacific Northwest National Laboratory is operated by Battelle for the U.S. Department of Energy under Contract DE-AC05-76RL01830.

\end{acknowledgments}

\section*{Data and Code Availability}
All data described in this paper are presented in the Appendix. Additional information supporting the findings of this study is available from the corresponding author upon reasonable request.

\appendix

\section{Benchmark Target Functions}
\label{app:benchmark_targets}

\begin{figure}[htbp]
    \centering
    \includegraphics[width=0.95\linewidth]{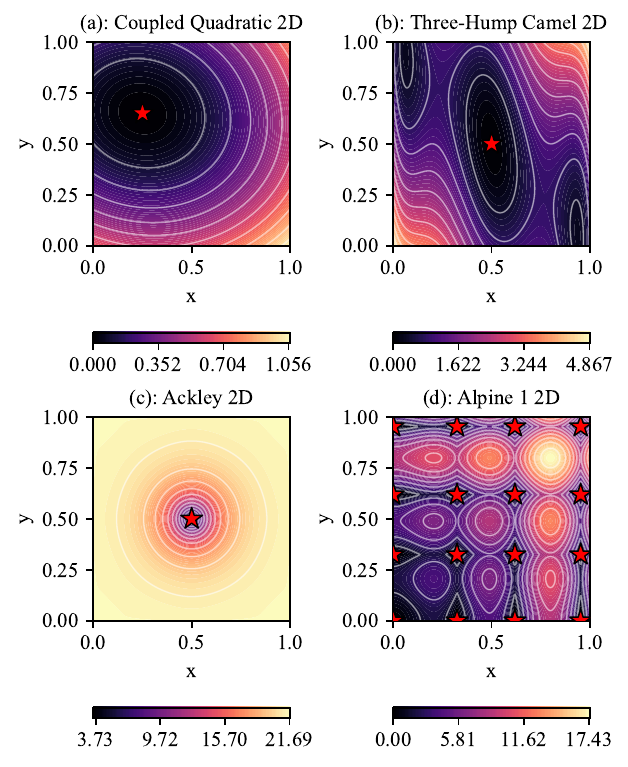}
    \caption{The four two-dimensional benchmark target functions and the corresponding locations of their global minima. The location of the global minima can be found in the main text.}
    \label{fig:benchmark_2d}
\end{figure}

This appendix lists the one- and two-dimensional benchmark target functions used in this work, together with their continuous global minimizers. The examples are adapted, with modifications, from the benchmark problems provided in the QHD GitHub repository~\cite{qhdGithub}. Following those benchmarks, all target functions are defined on normalized domains: one-dimensional targets use $u\in[0,1]$, and two-dimensional targets use $(x,y)\in[0,1]^2$.

\subsection{One-Dimensional Targets}

We use four one-dimensional target functions as benchmark problems. These functions are plotted in Fig.~\ref{fig:verify_circ}e to Fig.~\ref{fig:verify_circ}h.

\paragraph{Centered Quadratic:}
\begin{equation}
    f(u)=\left(u-\frac{1}{2}\right)^2.
\end{equation}
The unique global minimizer is
\begin{equation}
    u^\star=\frac{1}{2},\qquad f(u^\star)=0.
\end{equation}

\paragraph{Double-Well:}
\begin{equation}
    f(u)=4(u-0.3)^2(u-0.7)^2+0.02(u-0.55).
\end{equation}
The global minimizer is
\begin{equation}
    u^\star\approx 0.285901,
    \qquad f(u^\star)\approx -5.1456\times 10^{-3}.
\end{equation}

\paragraph{Cosine Periodic:}
\begin{equation}
    f(u)=1-\cos\bigl(2\pi(u-0.3)\bigr).
\end{equation}
On $[0,1]$, the unique global minimizer is
\begin{equation}
    u^\star=0.3,
    \qquad f(u^\star)=0.
\end{equation}

\paragraph{Two-Mode Cosine:}
\begin{equation}
\begin{aligned}
    f(u)
    &=1.2-\cos\bigl(2\pi(u-0.3)\bigr)
    +0.25\cos\bigl(4\pi(u-0.3)\bigr) \\
    &\quad +0.1\sin\bigl(2\pi(u-0.3)\bigr).
\end{aligned}
\end{equation}
The global minimizer is
\begin{equation}
    u^\star\approx 0.209657,
    \qquad f(u^\star)\approx 0.408533.
\end{equation}

\subsection{Two-Dimensional Targets}
We mainly focus on four two-dimensional target functions for resource analysis and verification of the performance of the QHD algorithm. These four functions are plotted in Fig.~\ref{fig:benchmark_2d}.

\paragraph{Coupled Quadratic 2D.}
\begin{equation}
\begin{aligned}
    f(x,y)
    &=(x-0.25)^2+1.4(y-0.65)^2 \\
    &\quad +0.2(x-0.25)(y-0.65).
\end{aligned}
\end{equation}
The unique global minimizer is
\begin{equation}
    (x^\star,y^\star)=(0.25,0.65),
    \qquad f(x^\star,y^\star)=0.
\end{equation}

\paragraph{Three-Hump Camel 2D.}
With $X=4x-2$ and $Y=2y-1$, the normalized three-hump camel target is
\begin{equation}
    f(x,y)=2X^2-\frac{21}{20}X^4+\frac{X^6}{6}+XY+Y^2.
\end{equation}
The unique global minimizer is
\begin{equation}
    (x^\star,y^\star)=(0.5,0.5),
    \qquad f(x^\star,y^\star)=0.
\end{equation}

\paragraph{Ackley 2D.}
With $X=-32.768+65.536x$ and $Y=-32.768+65.536y$, the normalized Ackley target is
\begin{equation}
\begin{aligned}
    f(x,y)
    &=-20\exp\left(-0.2\sqrt{\frac{X^2+Y^2}{2}}\right) \\
    &\quad -\exp\left(\frac{\cos(2\pi X)+\cos(2\pi Y)}{2}\right)
    +20+\mathrm{e}.
\end{aligned}
\end{equation}
The unique global minimizer is
\begin{equation}
    (x^\star,y^\star)=(0.5,0.5),
    \qquad f(x^\star,y^\star)=0.
\end{equation}

\paragraph{Alpine 1 2D.}
With $X=10x$ and $Y=10y$, the normalized Alpine 1 target is
\begin{equation}
    f(x,y)=\left|X\sin X+0.1X\right|+\left|Y\sin Y+0.1Y\right|.
\end{equation}
This target has multiple global minimizers. Defining
\begin{equation}
    \mathcal{S}=\{0,\ 0.324176,\ 0.618302,\ 0.952495\},
\end{equation}
the global minimizers are
\begin{equation}
    (x^\star,y^\star)\in\mathcal{S}\times\mathcal{S},
    \qquad f(x^\star,y^\star)=0.
\end{equation}

\begin{figure}[ht]
    \centering
    \includegraphics[width=0.75\linewidth]{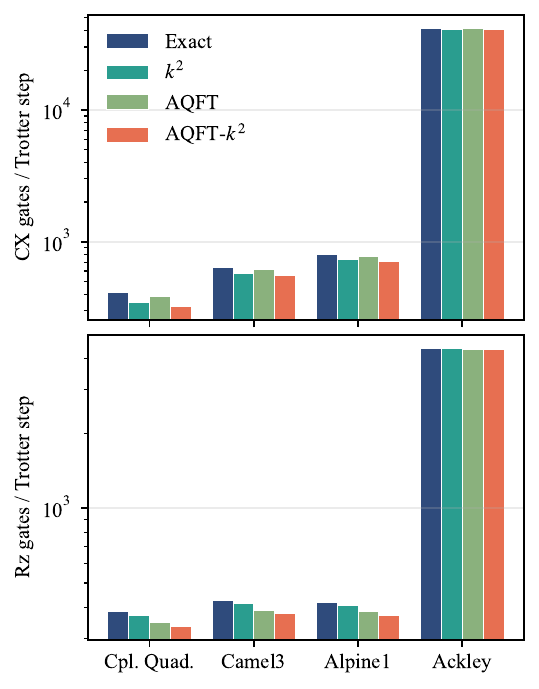}
    \caption{The CX and $R_z$ gate counts for four 2D benchmark problems using different kinetic-evolution approximations with resolution $N=64$.}
    \label{fig:all4_kinetic_cts}
\end{figure}

\section{Additional data for gate count scaling for kinetic evolutions} \label{app:gate_scaling}

\begin{figure*}[htbp]
    \centering
    \subfloat[Three-Hump Camel]{\includegraphics[width=0.32\textwidth]{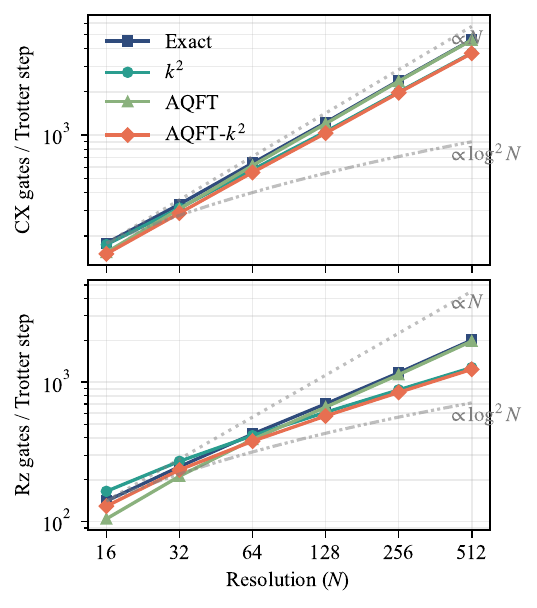}} \,
    \subfloat[Alpine1]{\includegraphics[width=0.32\textwidth]{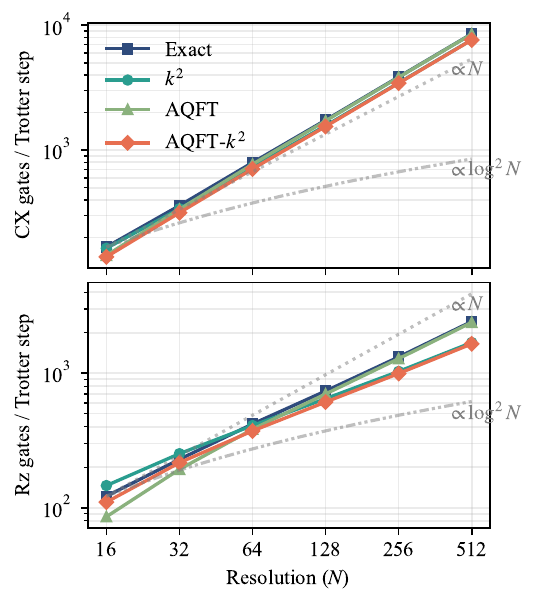}} \,
    \subfloat[Ackley]{\includegraphics[width=0.32\textwidth]{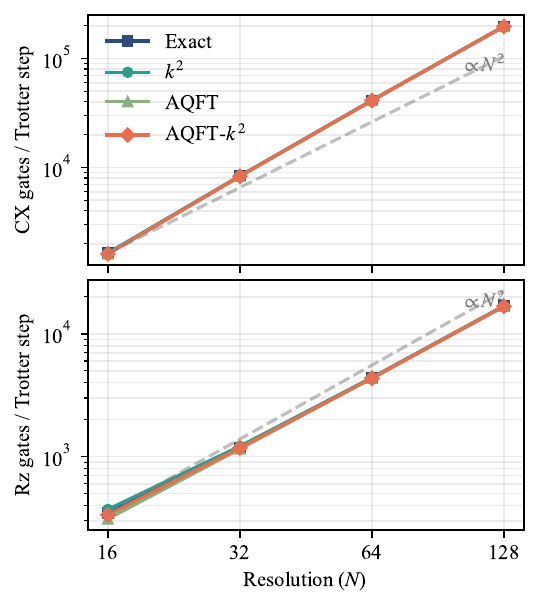}}
    \caption{Gate-count scaling for different kinetic-evolution approximations for (a) Three-Hump Camel, (b) Alpine 1, and (c) Ackley target functions. The settings are identical to the analysis that is reported in Fig.~\ref{fig:cq_kinetic}.}
    \label{fig:gate_count_scaling_app}
\end{figure*}

\begin{figure*}[htbp]
    \centering
    \includegraphics[width=0.9\textwidth]{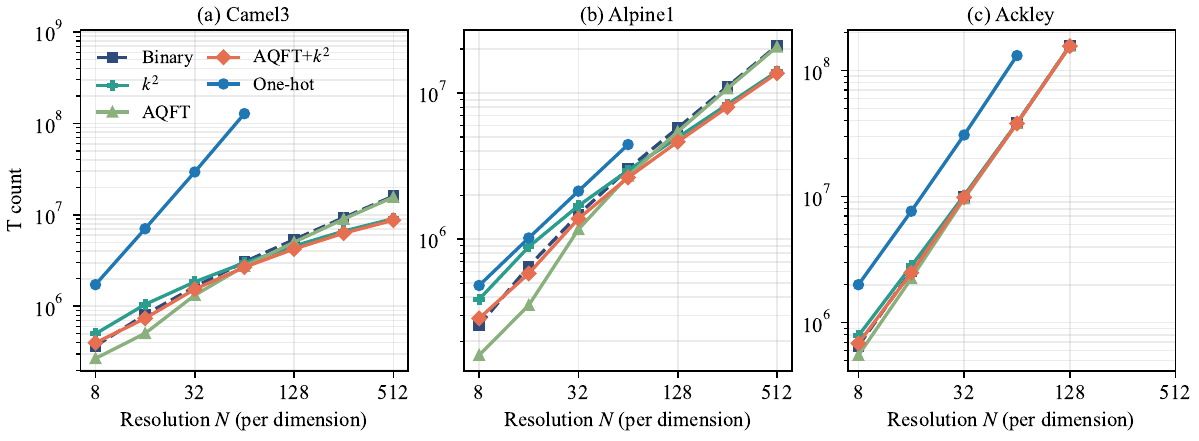}
    \caption{$T$-gate counts for (a) Three-Hump Camel, (b) Alpine 1, and (c) Ackley benchmark target functions. The curves compare the exact binary-encoding realization with variants using the low-momentum kinetic-spectrum approximation, the approximate-QFT subroutine, and their combination. The settings are identical to those used in Fig.~\ref{fig:t_gate}.}
    \label{fig:t_gate_scaling_app}
\end{figure*}

This section provides additional data and discussion for the gate-count scaling benchmarks. In Fig.~\ref{fig:all4_kinetic_cts}, we show the CX and $R_z$ gate counts for all four 2D benchmark problems. Because the same resolution and Trotter-step settings are used for all benchmarks, the kinetic-evolution circuit is the same across these cases. Therefore, the differences in gate counts arise solely from the potential evolution. In particular, the Ackley benchmark is dominated by the potential realization, as discussed in Sec.~\ref{sec:resource}.

In Fig.~\ref{fig:gate_count_scaling_app}, we plot the gate-count scaling for the remaining three 2D benchmark problems: Three-Hump Camel, Alpine 1, and Ackley. We use the same methods and configurations as in Fig.~\ref{fig:cq_kinetic}. Similar to the Coupled Quadratic case reported in the main text, the Three-Hump Camel benchmark shows a clear gate-count reduction when the low-momentum approximation is applied to the kinetic term, and the $R_z$ gate count approaches the expected $\log^2(N)$ scaling [see Fig.~\ref{fig:gate_count_scaling_app}a]. In contrast, the reductions are less pronounced for Alpine 1 and Ackley. For Ackley in particular, the reduction is marginal, and the scaling changes little because the total cost is dominated by the potential realization.

In Fig.~\ref{fig:t_gate_scaling_app}, we show the corresponding fault-tolerant $T$-gate count scaling for the Three-Hump Camel, Alpine 1, and Ackley benchmarks under exact and approximate binary kinetic realizations. These results are consistent with the coupled-quadratic analysis in the main text. The kinetic approximations provide larger savings when potential synthesis is not dominant, while Ackley shows smaller reductions because dense potential synthesis controls the total $T$ count.

\bibliography{ref}

@article{wu2025quantumdual,
  title={A quantum dual logarithmic barrier method for linear optimization},
  author={Wu, Zeguan and Sampourmahani, Pouya and Mohammadisiahroudi, Mohammadhossein and Terlaky, Tam{\'a}s},
  journal={INFORMS Journal on Optimization},
  year={2025},
  publisher={INFORMS}
}

@book{nielsen2010quantum,
  title={Quantum Computation and Quantum Information},
  author={Nielsen, Michael A and Chuang, Isaac L},
  year={2010},
  publisher={Cambridge University Press}
}

@inproceedings{shor1994algorithms,
  title={Algorithms for quantum computation: discrete logarithms and factoring},
  author={Shor, Peter W},
  booktitle={Proceedings 35th Annual Symposium on Foundations of Computer Science},
  pages={124--134},
  year={1994},
  organization={IEEE}
}

@article{abbas2024challenges,
  title={Challenges and opportunities in quantum optimization},
  author={Abbas, Amira and Ambainis, Andris and Augustino, Brandon and B{\"a}rtschi, Andreas and Buhrman, Harry and Coffrin, Carleton and Cortiana, Giorgio and Dunjko, Vedran and Egger, Daniel J and Elmegreen, Bruce G and others},
  journal={Nature Reviews Physics},
  volume={6},
  number={12},
  pages={718--735},
  year={2024},
  publisher={Nature Publishing Group}
}

@misc{qhdGithub,
  author={{Leng, Jiaqi and Hickman, Ethan and Li, Joseph and Wu, Xiaodi}},
  title={{Quantum Hamiltonian Descent}: numerical simulation, real-machine deployment, and benchmarking},
  howpublished={\url{https://github.com/jiaqileng/quantum-hamiltonian-descent}},
  note={Accessed: June 14, 2026}
}

@article{leng2023quantum,
  title={Quantum hamiltonian descent},
  author={Leng, Jiaqi and Hickman, Ethan and Li, Joseph and Wu, Xiaodi},
  journal={arXiv preprint arXiv:2303.01471},
  year={2023}
}

@article{kushnir2025qhdopt,
  title={QHDOPT: A software for nonlinear optimization with quantum hamiltonian descent},
  author={Kushnir, Samuel and Leng, Jiaqi and Peng, Yuxiang and Fan, Lei and Wu, Xiaodi},
  journal={INFORMS Journal on Computing},
  volume={37},
  number={1},
  pages={107--124},
  year={2025},
  publisher={INFORMS}
}

@article{leng2025hamiltonianembedding,
  title={Expanding hardware-efficiently manipulable Hilbert space via Hamiltonian embedding},
  author={Leng, Jiaqi and Li, Joseph and Peng, Yuxiang and Wu, Xiaodi},
  journal={Quantum},
  volume={9},
  pages={1857},
  year={2025},
  doi={10.22331/q-2025-09-11-1857},
  url={https://quantum-journal.org/papers/q-2025-09-11-1857/}
}

@article{an2021timeDependentUnbounded,
  title={Time-dependent unbounded {H}amiltonian simulation with vector norm scaling},
  author={An, Dong and Fang, Di and Lin, Lin},
  journal={Quantum},
  volume={5},
  pages={459},
  year={2021},
  publisher={Verein zur F{\"o}rderung des Open Access Publizierens in den Quantenwissenschaften}
}

@article{childs2021theory,
  title={Theory of {T}rotter error with commutator scaling},
  author={Childs, Andrew M and Su, Yuan and Tran, Minh C and Wiebe, Nathan and Zhu, Shuchen},
  journal={Physical Review X},
  volume={11},
  number={1},
  pages={011020},
  year={2021},
  publisher={APS}
}

@book{nocedal1999numerical,
  title={Numerical Optimization},
  author={Nocedal, Jorge and Wright, Stephen J},
  year={2006},
  publisher={Springer New York, NY}
}

@misc{kushnir2024qhdopt,
      title={QHDOPT: A Software for Nonlinear Optimization with Quantum Hamiltonian Descent},
      author={Samuel Kushnir and Jiaqi Leng and Yuxiang Peng and Lei Fan and Xiaodi Wu},
      year={2024},
      eprint={2409.03121},
      archivePrefix={arXiv},
      primaryClass={quant-ph},
      url={https://arxiv.org/abs/2409.03121},
}

@article{shor1999polynomial,
  title={Polynomial-time algorithms for prime factorization and discrete logarithms on a quantum computer},
  author={Shor, Peter W},
  journal={SIAM review},
  volume={41},
  number={2},
  pages={303--332},
  year={1999},
  publisher={SIAM}
}

@article{georgescu2014quantum,
  title={Quantum simulation},
  author={Georgescu, Iulia M and Ashhab, Sahel and Nori, Franco},
  journal={Reviews of Modern Physics},
  volume={86},
  number={1},
  pages={153},
  year={2014},
  publisher = {American Physical Society},
  doi = {10.1103/RevModPhys.86.153},
  url = {https://link.aps.org/doi/10.1103/RevModPhys.86.153}
}

@article{kandala2017hardware,
  title={Hardware-efficient variational quantum eigensolver for small molecules and quantum magnets},
  author={Kandala, Abhinav and Mezzacapo, Antonio and Temme, Kristan and Takita, Maika and Brink, Markus and Chow, Jerry M and Gambetta, Jay M},
  journal={Nature},
  volume={549},
  number={7671},
  pages={242--246},
  year={2017},
  publisher={Nature Publishing Group},
  doi={10.1038/nature23879},
  url={https://www.nature.com/articles/nature23879#citeas}
}

@article{rebentrost2018quantum,
  title={Quantum computational finance: Monte Carlo pricing of financial derivatives},
  author={Rebentrost, Patrick and Gupt, Brajesh and Bromley, Thomas R},
  journal={Physical Review A},
  volume={98},
  number={2},
  pages={022321},
  year={2018},
  publisher = {American Physical Society},
  doi = {10.1103/PhysRevA.98.022321},
  url = {https://link.aps.org/doi/10.1103/PhysRevA.98.022321}
}

@article{woerner2019quantum,
  title={Quantum risk analysis},
  author={Woerner, Stefan and Egger, Daniel J},
  journal={npj Quantum Information},
  volume={5},
  number={1},
  pages={1--8},
  year={2019},
  publisher={Nature Publishing Group},
  doi={10.1038/s41534-019-0130-6},
  url={https://www.nature.com/articles/s41534-019-0130-6}
}

@article{Das2008_annealing,
  title = {Colloquium: Quantum annealing and analog quantum computation},
  author = {Das, Arnab and Chakrabarti, Bikas K.},
  journal = {Rev. Mod. Phys.},
  volume = {80},
  issue = {3},
  pages = {1061--1081},
  numpages = {0},
  year = {2008},
  month = {Sep},
  publisher = {American Physical Society},
  doi = {10.1103/RevModPhys.80.1061},
  url = {https://link.aps.org/doi/10.1103/RevModPhys.80.1061}
}

@article{Tasseff2024_annealing,
author = {Tasseff, Byron and Albash, Tameem and Morrell, Zachary and Vuffray, Marc and Lokhov, Andrey Y. and Misra, Sidhant and Coffrin, Carleton},
title = {On the emerging potential of quantum annealing hardware for combinatorial optimization},
year = {2024},
issue_date = {Dec 2024},
publisher = {Kluwer Academic Publishers},
address = {USA},
volume = {30},
number = {5–6},
issn = {1381-1231},
url = {https://doi.org/10.1007/s10732-024-09530-5},
doi = {10.1007/s10732-024-09530-5},
journal = {Journal of Heuristics},
month = aug,
pages = {325–358},
numpages = {34}
}

@article{Zhou2020_qaoa,
  title = {Quantum Approximate Optimization Algorithm: Performance, Mechanism, and Implementation on Near-Term Devices},
  author = {Zhou, Leo and Wang, Sheng-Tao and Choi, Soonwon and Pichler, Hannes and Lukin, Mikhail D.},
  journal = {Phys. Rev. X},
  volume = {10},
  issue = {2},
  pages = {021067},
  numpages = {23},
  year = {2020},
  month = {Jun},
  publisher = {American Physical Society},
  doi = {10.1103/PhysRevX.10.021067},
  url = {https://link.aps.org/doi/10.1103/PhysRevX.10.021067}
}

@article{kim2025quantum,
  title={Quantum annealing for combinatorial optimization: a benchmarking study},
  author={Kim, Seongmin and Ahn, Sang-Woo and Suh, In-Saeng and Dowling, Alexander W. and Lee, Eungkyu and Luo, Tengfei},
  journal={npj Quantum Information},
  volume={11},
  number={77},
  year={2025},
  publisher={Nature Publishing Group},
  doi={10.1038/s41534-025-01020-1},
  url={https://www.nature.com/articles/s41534-025-01020-1}
}

@article{Eastin2009,
  title = {Restrictions on Transversal Encoded Quantum Gate Sets},
  author = {Eastin, Bryan and Knill, Emanuel},
  journal = {Phys. Rev. Lett.},
  volume = {102},
  issue = {11},
  pages = {110502},
  numpages = {4},
  year = {2009},
  month = {Mar},
  publisher = {American Physical Society},
  doi = {10.1103/PhysRevLett.102.110502},
  url = {https://link.aps.org/doi/10.1103/PhysRevLett.102.110502}
}

@article{McKay2017,
  title = {Efficient $Z$ gates for quantum computing},
  author = {McKay, David C. and Wood, Christopher J. and Sheldon, Sarah and Chow, Jerry M. and Gambetta, Jay M.},
  journal = {Phys. Rev. A},
  volume = {96},
  issue = {2},
  pages = {022330},
  numpages = {8},
  year = {2017},
  month = {Aug},
  publisher = {American Physical Society},
  doi = {10.1103/PhysRevA.96.022330},
  url = {https://link.aps.org/doi/10.1103/PhysRevA.96.022330}
}

@misc{gottesman1998,
      title={The Heisenberg Representation of Quantum Computers},
      author={Daniel Gottesman},
      year={1998},
      eprint={quant-ph/9807006},
      archivePrefix={arXiv},
      primaryClass={quant-ph},
      url={https://arxiv.org/abs/quant-ph/9807006},
}

@article{Aaronson2004,
  title = {Improved simulation of stabilizer circuits},
  author = {Aaronson, Scott and Gottesman, Daniel},
  journal = {Phys. Rev. A},
  volume = {70},
  issue = {5},
  pages = {052328},
  numpages = {14},
  year = {2004},
  month = {Nov},
  publisher = {American Physical Society},
  doi = {10.1103/PhysRevA.70.052328},
  url = {https://link.aps.org/doi/10.1103/PhysRevA.70.052328}
}

@article{Bombin2009,
        author = {H Bombin and M A Martin-Delgado},
        doi = {10.1088/1751-8113/42/9/095302},
        journal = {Journal of Physics A: Mathematical and Theoretical},
        month = {feb},
        number = {9},
        pages = {095302},
        title = {Quantum measurements and gates by code deformation},
        url = {https://dx.doi.org/10.1088/1751-8113/42/9/095302},
        volume = {42},
        year = {2009}}

@article{Anderson2014,
  title = {Fault-Tolerant Conversion between the Steane and Reed-Muller Quantum Codes},
  author = {Anderson, Jonas T. and Duclos-Cianci, Guillaume and Poulin, David},
  journal = {Phys. Rev. Lett.},
  volume = {113},
  issue = {8},
  pages = {080501},
  numpages = {5},
  year = {2014},
  month = {Aug},
  publisher = {American Physical Society},
  doi = {10.1103/PhysRevLett.113.080501},
  url = {https://link.aps.org/doi/10.1103/PhysRevLett.113.080501}
}

@article{Horsman2012,
        author = {Dominic Horsman and Austin G Fowler and Simon Devitt and Rodney Van Meter},
        doi = {10.1088/1367-2630/14/12/123011},
        journal = {New Journal of Physics},
        month = {dec},
        number = {12},
        pages = {123011},
        publisher = {IOP Publishing},
        title = {Surface code quantum computing by lattice surgery},
        url = {https://dx.doi.org/10.1088/1367-2630/14/12/123011},
        volume = {14},
        year = {2012}}

@article{Brown2017,
  title = {Poking Holes and Cutting Corners to Achieve Clifford Gates with the Surface Code},
  author = {Brown, Benjamin J. and Laubscher, Katharina and Kesselring, Markus S. and Wootton, James R.},
  journal = {Phys. Rev. X},
  volume = {7},
  issue = {2},
  pages = {021029},
  numpages = {20},
  year = {2017},
  month = {May},
  publisher = {American Physical Society},
  doi = {10.1103/PhysRevX.7.021029},
  url = {https://link.aps.org/doi/10.1103/PhysRevX.7.021029}
}

@article{Litinski2019,
  doi = {10.22331/q-2019-03-05-128},
  url = {https://doi.org/10.22331/q-2019-03-05-128},
  title = {A {G}ame of {S}urface {C}odes: {L}arge-{S}cale {Q}uantum {C}omputing with {L}attice {S}urgery},
  author = {Litinski, Daniel},
  journal = {{Quantum}},
  issn = {2521-327X},
  publisher = {{Verein zur F{\"{o}}rderung des Open Access Publizierens in den Quantenwissenschaften}},
  volume = {3},
  pages = {128},
  month = mar,
  year = {2019}
}

@article{Laflamme2014,
  title = {Using Concatenated Quantum Codes for Universal Fault-Tolerant Quantum Gates},
  author = {Jochym-O'Connor, Tomas and Laflamme, Raymond},
  journal = {Phys. Rev. Lett.},
  volume = {112},
  issue = {1},
  pages = {010505},
  numpages = {5},
  year = {2014},
  month = {Jan},
  publisher = {American Physical Society},
  doi = {10.1103/PhysRevLett.112.010505},
  url = {https://link.aps.org/doi/10.1103/PhysRevLett.112.010505}
}

@misc{gidney2024cultivation,
      title={Magic state cultivation: growing T states as cheap as CNOT gates},
      author={Craig Gidney and Noah Shutty and Cody Jones},
      year={2024},
      eprint={2409.17595},
      archivePrefix={arXiv},
      primaryClass={quant-ph},
      url={https://arxiv.org/abs/2409.17595},
}

@misc{rosenfeld2025,
      title={Magic state cultivation on a superconducting quantum processor},
      author={Emma Rosenfeld and Craig Gidney and Gabrielle Roberts and Alexis Morvan and Nathan Lacroix and Dvir Kafri and Jeffrey Marshall and Ming Li and Volodymyr Sivak and Dmitry Abanin and Amira Abbas and Rajeev Acharya and Laleh Aghababaie Beni and Georg Aigeldinger and Ross Alcaraz and Sayra Alcaraz and Trond I. Andersen and Markus Ansmann and Frank Arute and Kunal Arya and Walt Askew and Nikita Astrakhantsev and Juan Atalaya and Ryan Babbush and Brian Ballard and Joseph C. Bardin and Hector Bates and Andreas Bengtsson and Majid Bigdeli Karimi and Alexander Bilmes and Simon Bilodeau and Felix Borjans and Jenna Bovaird and Dylan Bowers and Leon Brill and Peter Brooks and Michael Broughton and David A. Browne and Brett Buchea and Bob B. Buckley and Tim Burger and Brian Burkett and Nicholas Bushnell and Jamal Busnaina and Anthony Cabrera and Juan Campero and Hung-Shen Chang and Silas Chen and Zijun Chen and Ben Chiaro and Liang-Ying Chih and Agnetta Y. Cleland and Bryan Cochrane and Matt Cockrell and Josh Cogan and Paul Conner and Harold Cook and Rodrigo G. Cortiñas and William Courtney and Alexander L. Crook and Ben Curtin and Martin Damyanov and Sayan Das and Dripto M. Debroy and Sean Demura and Paul Donohoe and Ilya Drozdov and Andrew Dunsworth and Valerie Ehimhen and Alec Eickbusch and Aviv Moshe Elbag and Lior Ella and Mahmoud Elzouka and David Enriquez and Catherine Erickson and Lara Faoro and Vinicius S. Ferreira and Marcos Flores and Leslie Flores Burgos and Sam Fontes and Ebrahim Forati and Jeremiah Ford and Brooks Foxen and Masaya Fukami and Alan Wing Lun Fung and Lenny Fuste and Suhas Ganjam and Gonzalo Garcia and Christopher Garrick and Robert Gasca and Helge Gehring and Robert Geiger and Élie Genois and William Giang and Dar Gilboa and James E. Goeders and Edward C. Gonzales and Raja Gosula and Stijn J. de Graaf and Alejandro Grajales Dau and Dietrich Graumann and Joel Grebel and Alex Greene and Jonathan A. Gross and Jose Guerrero and Loïck Le Guevel and Tan Ha and Steve Habegger and Tanner Hadick and Ali Hadjikhani and Michael C. Hamilton and Monica Hansen and Matthew P. Harrigan and Sean D. Harrington and Jeanne Hartshorn and Stephen Heslin and Paula Heu and Oscar Higgott and Reno Hiltermann and Jeremy Hilton and Hsin-Yuan Huang and Mike Hucka and Christopher Hudspeth and Ashley Huff and William J. Huggins and Lev B. Ioffe and Evan Jeffrey and Shaun Jevons and Zhang Jiang and Xiaoxuan Jin and Chaitali Joshi and Pavol Juhas and Andreas Kabel and Hui Kang and Kiseo Kang and Amir H. Karamlou and Ryan Kaufman and Kostyantyn Kechedzhi and Tanuj Khattar and Mostafa Khezri and Seon Kim and Paul V. Klimov and Can M. Knaut and Bryce Kobrin and Alexander N. Korotkov and Fedor Kostritsa and John Mark Kreikebaum and Ryuho Kudo and Ben Kueffler and Arun Kumar and Vladislav D. Kurilovich and Vitali Kutsko and Tiano Lange-Dei and Brandon W. Langley and Pavel Laptev and Kim-Ming Lau and Emma Leavell and Justin Ledford and Joy Lee and Kenny Lee and Brian J. Lester and Wendy Leung and Lily Li and Wing Yan Li and Alexander T. Lill and William P. Livingston and Matthew T. Lloyd and Aditya Locharla and Laura De Lorenzo and Erik Lucero and Daniel Lundahl and Aaron Lunt and Sid Madhuk and Aniket Maiti and Ashley Maloney and Salvatore Mandrà and Leigh S. Martin and Orion Martin and Eric Mascot and Paul Masih Das and Dmitri Maslov and Melvin Mathews and Cameron Maxfield and Jarrod R. McClean and Matt McEwen and Seneca Meeks and Anthony Megrant and Kevin C. Miao and Zlatko K. Minev and Reza Molavi and Sebastian Molina and Shirin Montazeri and Charles Neill and Michael Newman and Anthony Nguyen and Murray Nguyen and Chia-Hung Ni and Murphy Yuezhen Niu and Nicholas Noll and Logan Oas and William D. Oliver and Raymond Orosco and Kristoffer Ottosson and Alice Pagano and Agustin Di Paolo and Sherman Peek and David Peterson and Alex Pizzuto and Elias Portoles and Rebecca Potter and Orion Pritchard and Michael Qian and Chris Quintana and Ganesh Ramachandran and Arpit Ranadive and Matthew J. Reagor and Rachel Resnick and David M. Rhodes and Daniel Riley and Roberto Rodriguez and Emma Ropes and Lucia B. De Rose and Eliott Rosenberg and Dario Rosenstock and Elizabeth Rossi and Pedram Roushan and David A. Rower and Robert Salazar and Kannan Sankaragomathi and Murat Can Sarihan and Max Schaefer and Sebastian Schroeder and Henry F. Schurkus and Aria Shahingohar and Michael J. Shearn and Aaron Shorter and Noah Shutty and Vladimir Shvarts and Spencer Small and W. Clarke Smith and David A. Sobel and Barrett Spells and Sofia Springer and George Sterling and Jordan Suchard and Aaron Szasz and Alexander Sztein and Madeline Taylor and Jothi Priyanka Thiruraman and Douglas Thor and Dogan Timucin and Eifu Tomita and Alfredo Torres and M. Mert Torunbalci and Hao Tran and Abeer Vaishnav and Justin Vargas and Sergey Vdovichev and Guifre Vidal and Benjamin Villalonga and Catherine Vollgraff Heidweiller and Meghan Voorhees and Steven Waltman and Jonathan Waltz and Shannon X. Wang and Danni Wang and Brayden Ware and James D. Watson and Yonghua Wei and Travis Weidel and Theodore White and Kristi Wong and Bryan W. K. Woo and Christopher J. Wood and Maddy Woodson and Cheng Xing and Z. Jamie Yao and Ping Yeh and Bicheng Ying and Juhwan Yoo and Noureldin Yosri and Elliot Young and Grayson Young and Adam Zalcman and Ran Zhang and Yaxing Zhang and Ningfeng Zhu and Nicholas Zobrist and Zhenjie Zou and Hartmut Neven and Sergio Boixo and Cody Jones and Julian Kelly and Alexandre Bourassa and Kevin J. Satzinger},
      year={2025},
      eprint={2512.13908},
      archivePrefix={arXiv},
      primaryClass={quant-ph},
      url={https://arxiv.org/abs/2512.13908},
}

@article{litinski2019magic,
  title={Magic state distillation: Not as costly as you think},
  author={Litinski, Daniel},
  journal={Quantum},
  volume={3},
  pages={205},
  year={2019},
  publisher={Verein zur F{\"o}rderung des Open Access Publizierens in den Quantenwissenschaften}
}

@article{ross2014optimal,
  title={Optimal ancilla-free Clifford+ T approximation of z-rotations},
  author={Ross, Neil J and Selinger, Peter},
  journal={arXiv preprint arXiv:1403.2975},
  year={2014}
}

@article{aravena2023recent,
  title={Recent Developments in Security-Constrained {AC} Optimal Power Flow: Overview of Challenge 1 in the {ARPA-E} Grid Optimization Competition},
  author={Aravena, Ignacio and Molzahn, Daniel K and Zhang, Shixuan and Petra, Cosmin G and Curtis, Frank E and Tu, Shenyinying and W{\"a}chter, Andreas and Wei, Ermin and Wong, Elizabeth and Gholami, Amin and Sun, Kaizhao and Sun, Xu Andy and Elbert, Stephen T and Holzer, Jesse T and Veeramany, Arun},
  journal={Operations Research},
  volume={71},
  number={6},
  pages={1997--2014},
  year={2023},
  publisher={INFORMS},
  doi={10.1287/opre.2022.0315},
  url={https://doi.org/10.1287/opre.2022.0315}
}

@techreport{holzer2024grid,
  title={Grid Optimization Competition Challenge 3 Problem Formulation},
  author={Holzer, Jesse and Coffrin, Carleton and DeMarco, Christopher and Duthu, Ray and Elbert, Stephen and Eldridge, Brent and Elgindy, Tarek and Garcia, Manuel and Greene, Scott and Guo, Nongchao and Hale, Elaine and Lesieutre, Bernard and Mak, Terrence and McMillan, Colin and Mittelmann, Hans and Oh, Hyungseon and O'Neill, Richard and Overbye, Thomas and Palmintier, Bryan and Parker, Robbert and Safdarian, Farnaz and Tbaileh, Ahmad and Van Hentenryck, Pascal and Veeramany, Arun and Wangen, Steve and Wert, Jessica},
  institution={Pacific Northwest National Laboratory},
  year={2024},
  doi={10.2172/2337844},
  url={https://doi.org/10.2172/2337844}
}

@article{low2014convex,
  title={Convex Relaxation of Optimal Power Flow---Part {I}: Formulations and Equivalence},
  author={Low, Steven H},
  journal={IEEE Transactions on Control of Network Systems},
  volume={1},
  number={1},
  pages={15--27},
  year={2014},
  publisher={IEEE},
  doi={10.1109/TCNS.2014.2309732},
  url={https://doi.org/10.1109/TCNS.2014.2309732}
}

@article{markowitz1952portfolio,
  title={Portfolio Selection},
  author={Markowitz, Harry},
  journal={The Journal of Finance},
  volume={7},
  number={1},
  pages={77--91},
  year={1952},
  doi={10.1111/j.1540-6261.1952.tb01525.x},
  url={https://doi.org/10.1111/j.1540-6261.1952.tb01525.x}
}

@book{cornuejols2007optimization,
  title={Optimization Methods in Finance},
  author={Cornu{\'e}jols, G{\'e}rard and T{\"u}t{\"u}nc{\"u}, Reha},
  year={2007},
  publisher={Cambridge University Press},
  doi={10.1017/CBO9780511753886},
  url={https://doi.org/10.1017/CBO9780511753886}
}

@book{lavalle2006planning,
  title={Planning Algorithms},
  author={LaValle, Steven M},
  year={2006},
  publisher={Cambridge University Press},
  doi={10.1017/CBO9780511546877},
  url={https://doi.org/10.1017/CBO9780511546877}
}

@book{betts2010practical,
  title={Practical Methods for Optimal Control and Estimation Using Nonlinear Programming},
  author={Betts, John T},
  edition={2},
  year={2010},
  publisher={SIAM},
  doi={10.1137/1.9780898718577},
  url={https://doi.org/10.1137/1.9780898718577}
}

@inproceedings{zafar2017fairness,
  title={Fairness Constraints: Mechanisms for Fair Classification},
  author={Zafar, Muhammad Bilal and Valera, Isabel and Rodriguez, Manuel Gomez and Gummadi, Krishna P},
  booktitle={Proceedings of the 20th International Conference on Artificial Intelligence and Statistics},
  series={Proceedings of Machine Learning Research},
  volume={54},
  pages={962--970},
  year={2017},
  publisher={PMLR},
  url={https://proceedings.mlr.press/v54/zafar17a.html}
}

@article{tibshirani1996regression,
  title={Regression Shrinkage and Selection via the Lasso},
  author={Tibshirani, Robert},
  journal={Journal of the Royal Statistical Society: Series B (Methodological)},
  volume={58},
  number={1},
  pages={267--288},
  year={1996},
  doi={10.1111/j.2517-6161.1996.tb02080.x},
  url={https://doi.org/10.1111/j.2517-6161.1996.tb02080.x}
}

@article{xu2009robustness,
  title={Robustness and Regularization of Support Vector Machines},
  author={Xu, Huan and Caramanis, Constantine and Mannor, Shie},
  journal={Journal of Machine Learning Research},
  volume={10},
  number={51},
  pages={1485--1510},
  year={2009},
  url={https://jmlr.org/papers/v10/xu09b.html}
}

@software{nwqec_toolkit,
  author = {Wang, Meng and Liu, Chenxu and Li, Ang and Aniceto, Allister},
  title = {NWQEC: A toolkit for fault-tolerant quantum circuit transpilation and T-count optimization},
  url = {https://github.com/pnnl/nwqec},
  version = {0.1.1},
  year = {2025}
}

@article{wang2024optimizing,
  title={Optimizing FTQC Programs through QEC Transpiler and Architecture Codesign},
  author={Wang, Meng and Liu, Chenxu and Stein, Samuel and Ding, Yufei and Das, Poulami and Nair, Prashant J and Li, Ang},
  journal={arXiv preprint arXiv:2412.15434},
  year={2024}
}

@article{wang2025tableau,
  title={Tableau-Based Framework for Efficient Logical Quantum Compilation},
  author={Wang, Meng and Liu, Chenxu and Garner, Sean and Stein, Samuel and Ding, Yufei and Nair, Prashant J and Li, Ang},
  journal={arXiv preprint arXiv:2509.02721},
  year={2025}
}

@unpublished{song2026nwqre,
  title={{NWQRE}: A Modular Quantum Resource Estimation Workflow for Fault-Tolerant Algorithms},
  author={Song, Zhixin and Wang, Meng and Zheng, Muqing and Stein, Samuel and Bryngelson, Spencer H. and M{\"u}lmenst{\"a}dt, Johannes and Li, Xiangyu and Li, Ang and Liu, Chenxu},
  note={Manuscript in preparation},
  year={2026}
}

@misc{wu2026ALQHD,
      title={Benchmarking and Resource Analysis for Augmented-Lagrangian Quantum Hamiltonian Descent}, 
      author={Zeguan Wu and Mingze Li and Muqing Zheng and Meng Wang and Junyu Liu and Samuel Stein and Ang Li and Yousu Chen and Chenxu Liu},
      year={2026},
      eprint={2605.12066},
      archivePrefix={arXiv},
      primaryClass={quant-ph},
      url={https://arxiv.org/abs/2605.12066}, 
}

@misc{li2025QHD,
      title={Quantum Hamiltonian Descent based Augmented Lagrangian Method for Constrained Nonconvex Nonlinear Optimization}, 
      author={Mingze Li and Lei Fan and Zhu Han},
      year={2025},
      eprint={2508.02969},
      archivePrefix={arXiv},
      primaryClass={math.OC},
      url={https://arxiv.org/abs/2508.02969}, 
}

@misc{klappenecker2001QDST,
      title={On the Irresistible Efficiency of Signal Processing Methods in Quantum Computing}, 
      author={Andreas Klappenecker and Martin Roetteler},
      year={2001},
      eprint={quant-ph/0111039},
      archivePrefix={arXiv},
      primaryClass={quant-ph},
      url={https://arxiv.org/abs/quant-ph/0111039}, 
}

@misc{ahmadkhaniha2025qrtliblibraryfastquantum,
      title={QRTlib: A Library for Fast Quantum Real Transforms}, 
      author={Armin Ahmadkhaniha and Lu Chen and Jake Doliskani and Zhifu Sun},
      year={2025},
      eprint={2510.16625},
      archivePrefix={arXiv},
      primaryClass={quant-ph},
      url={https://arxiv.org/abs/2510.16625}, 
}

@article{Bravyi2016,
  title = {Trading Classical and Quantum Computational Resources},
  author = {Bravyi, Sergey and Smith, Graeme and Smolin, John A.},
  journal = {Phys. Rev. X},
  volume = {6},
  issue = {2},
  pages = {021043},
  numpages = {14},
  year = {2016},
  month = {Jun},
  publisher = {American Physical Society},
  doi = {10.1103/PhysRevX.6.021043},
  url = {https://link.aps.org/doi/10.1103/PhysRevX.6.021043}
}

@article{Amy2017CNOTPhase,
	author = {Amy, Matthew and Azimzadeh, Parsiad and Mosca, Michele},
	doi = {10.1088/2058-9565/aad8ca},
	journal = {Quantum Science and Technology},
	month = {sep},
	number = {1},
	pages = {015002},
	publisher = {IOP Publishing},
	title = {On the controlled-NOT complexity of controlled-NOT--phase circuits},
	url = {https://doi.org/10.1088/2058-9565/aad8ca},
	volume = {4},
	year = {2018}}

@misc{Cowtan2019PhaseGadget,
  title={Phase Gadget Synthesis for Shallow Circuits},
  author={Cowtan, Alexander and Dilkes, Silas and Duncan, Ross and Simmons, Will and Sivarajah, Seyon},
  year={2019},
  eprint={1906.01734},
  archivePrefix={arXiv},
  primaryClass={quant-ph},
  doi={10.4204/EPTCS.318.13},
  url={https://arxiv.org/abs/1906.01734}
}

@misc{Vandaele2021PhasePolynomial,
  title={Phase polynomials synthesis algorithms for {NISQ} architectures and beyond},
  author={Vandaele, Vivien and Martiel, Simon and Goubault de Brugi{\`e}re, Timoth{\'e}e},
  year={2021},
  eprint={2104.00934},
  archivePrefix={arXiv},
  primaryClass={quant-ph},
  url={https://arxiv.org/abs/2104.00934}
}

@article{Chang2022GrayCode,
  title = {Improving {Schr\"odinger} Equation Implementations with Gray Code for Adiabatic Quantum Computers},
  author = {Chang, Chia Cheng and McElvain, Kenneth S. and Rrapaj, Ermal and Wu, Yantao},
  journal = {PRX Quantum},
  volume = {3},
  issue = {2},
  pages = {020356},
  year = {2022},
  month = {Jun},
  publisher = {American Physical Society},
  doi = {10.1103/PRXQuantum.3.020356},
  url = {https://journals.aps.org/prxquantum/abstract/10.1103/PRXQuantum.3.020356}
}

@article{Hu2024Schrodingerisation,
  title = {Quantum Circuits for Partial Differential Equations via {Schr\"odingerisation}},
  author = {Hu, Junpeng and Jin, Shi and Liu, Nana and Zhang, Lei},
  journal = {Quantum},
  volume = {8},
  pages = {1563},
  year = {2024},
  month = {Dec},
  doi = {10.22331/q-2024-12-12-1563},
  url = {https://quantum-journal.org/papers/q-2024-12-12-1563/}
}

@article{Hadfield2021,
author = {Hadfield, Stuart},
title = {On the Representation of Boolean and Real Functions as Hamiltonians for Quantum Computing},
year = {2021},
issue_date = {December 2021},
publisher = {Association for Computing Machinery},
address = {New York, NY, USA},
volume = {2},
number = {4},
url = {https://doi.org/10.1145/3478519},
doi = {10.1145/3478519},
journal = {ACM Transactions on Quantum Computing},
month = dec,
articleno = {18},
numpages = {21}
}

@article{Welch_2014,
	author = {Welch, Jonathan and Greenbaum, Daniel and Mostame, Sarah and Aspuru-Guzik, Alan},
	doi = {10.1088/1367-2630/16/3/033040},
	journal = {New Journal of Physics},
	month = {mar},
	number = {3},
	pages = {033040},
	publisher = {IOP Publishing},
	title = {Efficient quantum circuits for diagonal unitaries without ancillas},
	url = {https://doi.org/10.1088/1367-2630/16/3/033040},
	volume = {16},
	year = {2014}}

@article{Zhuang2024Depth,
  title = {Depth-optimized quantum circuit synthesis for diagonal unitary operators with asymptotically optimal gate count},
  author = {Zhang, Shihao and Huang, Kai and Li, Lvzhou},
  journal = {Phys. Rev. A},
  volume = {109},
  issue = {4},
  pages = {042601},
  numpages = {11},
  year = {2024},
  month = {Apr},
  publisher = {American Physical Society},
  doi = {10.1103/PhysRevA.109.042601},
  url = {https://link.aps.org/doi/10.1103/PhysRevA.109.042601}
}

@article{Nam2020AQFT,
  title = {Approximate quantum Fourier transform with {O}(n log(n)) {T} gates},
  author = {Nam, Yunseong and Su, Yuan and Maslov, Dmitri},
  journal = {npj Quantum Information},
  volume = {6},
  number = {1},
  pages = {26},
  year = {2020},
  doi = {10.1038/s41534-020-0257-5},
  url = {https://doi.org/10.1038/s41534-020-0257-5}
}

@article{NISQ_Preskill,
  doi = {10.22331/q-2018-08-06-79},
  url = {https://doi.org/10.22331/q-2018-08-06-79},
  title = {Quantum {C}omputing in the {NISQ} era and beyond},
  author = {Preskill, John},
  journal = {{Quantum}},
  issn = {2521-327X},
  publisher = {{Verein zur F{\"{o}}rderung des Open Access Publizierens in den Quantenwissenschaften}},
  volume = {2},
  pages = {79},
  month = aug,
  year = {2018}
}

@article{Megaquop,
author = {Preskill, John},
title = {Beyond NISQ: The Megaquop Machine},
year = {2025},
issue_date = {September 2025},
publisher = {Association for Computing Machinery},
address = {New York, NY, USA},
volume = {6},
number = {3},
url = {https://doi.org/10.1145/3723153},
doi = {10.1145/3723153},
journal = {ACM Transactions on Quantum Computing},
month = apr,
articleno = {18},
numpages = {7}
}

\end{document}